\documentclass[final,5p,times]{elsarticle}

\usepackage{amssymb}

\usepackage[utf8]{inputenc}

\usepackage{listings} 
\usepackage{graphicx}
\usepackage{amsmath}
\usepackage{float} 
\usepackage[T1]{fontenc} 
\usepackage{courier} 
\usepackage{verbatim} 
\usepackage{wrapfig}
\usepackage{xcolor} 
\usepackage{tcolorbox} 
\usepackage{tabularx} 
\usepackage{array} 
\usepackage{colortbl} 
\tcbuselibrary{skins}
\usepackage{rotating} 
\usepackage{multicol} 
\usepackage{subfigure} 
\usepackage{booktabs} 
\usepackage{multirow} 
\usepackage{enumitem} 
\usepackage[section]{placeins} 
\usepackage{hyperref} 
\usepackage{pdfpages} 
\usepackage[normalem]{ulem} 
\usepackage{amssymb}
\usepackage{pifont}
\usepackage{adjustbox} 
\usepackage{url} 
\usepackage{lineno} 
\usepackage[]{algorithm2e} 
\usepackage{lmodern} 
\usepackage{mathtools} 
\usepackage{tabularx} 
\usepackage[htt]{hyphenat} 

\lstdefinelanguage{Markdown}{
	morekeywords={},
	basicstyle=\small\ttfamily, 
	captionpos=b, 
	extendedchars=true, 
	tabsize=2, 
	columns=fixed, 
	keepspaces=true, 
	showstringspaces=false, 
	breaklines=true, 
	frame=trbl, 
	frameround=tttt, 
	framesep=4pt, 
	numbers=left, 
	numberstyle=\tiny\ttfamily, 
	commentstyle=\color{eclipseGreen}, 
	keywordstyle=\color{eclipsePurple}, 
	stringstyle=\color{eclipseBlue} 
}
		
\lstdefinelanguage{YAML}{
	morekeywords={true,false,null,y,n},
	basicstyle=\ttfamily\scriptsize, 
	captionpos=b, 
	extendedchars=true, 
	tabsize=1, 
	columns=fixed, 
	keepspaces=true, 
	showstringspaces=false, 
	breaklines=true, 
	frame=trbl, 
	frameround=tttt, 
	framesep=4pt, 
	numbers=left, 
	numberstyle=\tiny\ttfamily, 
	commentstyle=\color{eclipseGreen}, 
	keywordstyle=\color{darkgray}\bfseries, 
	stringstyle=\color{eclipseBlue},
    stepnumber=1,
    numbersep=-3pt,
    showspaces=false
}
		
\lstdefinelanguage{JSON}{
	morekeywords={true,false,null,y,n},
	basicstyle=\small\ttfamily, 
	captionpos=b, 
	extendedchars=true, 
	tabsize=2, 
	columns=fixed, 
	keepspaces=true, 
	showstringspaces=false, 
	breaklines=true, 
	frame=trbl, 
	frameround=tttt, 
	framesep=4pt, 
	numbers=left, 
	numberstyle=\tiny\ttfamily, 
	commentstyle=\color{eclipseGreen}, 
	keywordstyle=\color{darkgray}\bfseries, 
	stringstyle=\color{eclipseBlue}
}
		
\lstdefinelanguage{XML}{
	morekeywords={true,false,null,y,n},
	basicstyle=\small\ttfamily, 
	captionpos=b, 
	extendedchars=true, 
	tabsize=2, 
	columns=fixed, 
	keepspaces=true, 
	showstringspaces=false, 
	breaklines=true, 
	frame=trbl, 
	frameround=tttt, 
	framesep=4pt, 
	numbers=left, 
	numberstyle=\tiny\ttfamily, 
	commentstyle=\color{eclipseGreen}, 
	keywordstyle=\color{darkgray}\bfseries, 
	stringstyle=\color{eclipseBlue}
}
\usepackage[normalem]{ulem} 
\usepackage{xspace} 
\usepackage{xcolor}
\definecolor{colgray}{gray}{0.75}

\usepackage{ifthen}
\newboolean{showcomments}
\setboolean{showcomments}{true} 
\ifthenelse{\boolean{showcomments}}
  {\newcommand{\nb}[2]{
    \fcolorbox{gray}{yellow}{\bfseries\sffamily\scriptsize#1}
    {$\blacktriangleright$#2$\blacktriangleleft$}
   }
   
  }
  {\newcommand{\nb}[2]{}
   
  }

\journal{Computer Standards \& Interfaces}

\begin{document}

\begin{frontmatter}



\title{Pricing4APIs: A Rigorous Model for RESTful API Pricings}


\author[1]{Rafael Fresno-Aranda\corref{cor1}}
\ead{rfresno@us.es}

\author[1]{Pablo Fernandez}
\ead{pablofm@us.es}

\author[2]{Antonio Gamez-Diaz}
\ead{agamez2@us.es}

\author[1]{Amador Duran}
\ead{amador@us.es}

\author[1]{Antonio Ruiz-Cortes}
\ead{aruiz@us.es}

\cortext[cor1]{Corresponding author}
\affiliation[1]{organization={SCORE Lab, I3US Institute, Universidad de Sevilla},
            addressline={Avda. Reina Mercedes S/N}, 
            city={Seville},
            postcode={41012},
            country={Spain}}
\affiliation[2]{organization={Independent Researcher},
                country={Spain}}

\begin{abstract}
		        
		
        APIs are increasingly becoming new business assets for organizations and consequently, API functionality and its pricing should be precisely defined for customers. Pricing is typically composed by different plans that specify a range of limitations, e.g., a Free plan allows 100 monthly requests while a Gold plan has 10000 requests per month. In this context, the OpenAPI Specification (OAS) has emerged to model the functional part of an API, becoming a de facto industry standard and boosting a rich ecosystem of vendor-neutral tools to assist API providers and consumers. In contrast, there is no proposal for modeling API pricings (i.e. their plans and limitations) and this lack hinders the creation of tools that can leverage this information. To deal with this gap, this paper presents a pricing modeling framework that includes: (a) \textit{Pricing4APIs} model, a comprehensive and rigorous model of API pricings, along \textit{SLA4OAI}, a serialization that extends OAS; (b) an operation to validate the description of API pricings, with a toolset (\textit{sla4oai-analyzer}) that has been developed to automate this operation. Additionally, we analyzed 268 real-world APIs to assess the expressiveness of our proposal and created a representative dataset of 54 pricing models to validate our framework.
		       
		        
		        
		        

\end{abstract}


\begin{highlights}
\item Pricing4APIs, a novel model for RESTful API pricings, and a set of validity criteria.
\item An extension of OpenAPI Specification to serialize Pricing4APIs.
\item A dataset of 54 real-world pricings modeled with our proposal.
\item Analysis of the expressiveness of Pricing4APIs using a systematic review of 268 APIs.
\item An automated validation tool to check for pricing inconsistencies.
\end{highlights}

\begin{keyword}
Web services \sep RESTful APIs \sep Pricings \sep Limitations \sep Quota \sep Rate



\end{keyword}

\end{frontmatter}




\section{Introduction}\label{sec:introduction}

Today, APIs are regarded as a new form of business product, and ever more organizations are publicly opening up access to their APIs as a way to create new business opportunities in this so-called API Economy. Indeed, this trend has been given a boost by the shift towards microservice architectures as the preferred choice for the construction of cloud-native Software as a Service. Since these architectures typically promote the deployment and integration of components (i.e., microservices) by means of RESTful Web APIs (henceforth, for the sake of simplicity, APIs), they pave the way for easy connection to external APIs (as service consumers) or opening up internal APIs to the market (as service providers). Moreover, in recent years, a successful effort has been made for standardization, the Open API Specification\footnote{More details can be found at \url{https://www.openapis.org}} (OAS) which was aimed at describing the functional part of APIs (i.e., the available resources and operations). This \textit{de-facto} standard has led to the creation of a rich open ecosystem of tools and techniques to help in the development and evolution of APIs and microservice architectures proposed by Academia (such as \cite{gonzalez2023improving} or \cite{martin2021specification}) or in extensive tool catalogs in the Industry (such as the
\url{https://openapi.tools/} with over 350 tools listed).

In this context, from a non-functional perspective, defining business models and plans with API limitations, such as quotas or rates, has become crucial for the regulation of the behavior expected from all participants and for a guarantee of a certain service level. For instance, a premium-tier Google Maps API user might expect a higher request rate, let us say 300 requests per second (req/s), compared to a basic-tier user limited to, say, 50 req/s.  However, information on API limitations is neither structured nor standardized, as shown in ~\cite{Gamez-Diaz2017_icsoc_main}. As a consequence, the ecosystem of tools cannot benefit from this information to support the process of developing, consuming, or operating APIs in tasks such as:
\begin{itemize}
    \item Estimated Capacity Analysis: it becomes paramount to accurately estimate the capacity of an API given its limitations. For instance, a developer or an API provider might want to calculate the maximum number of requests that can be made within a certain time frame given the API's tiered limitations. Knowing this capacity helps in proper planning the infrastructure allocation for the API providers and can also serve as an insightful metric for potential API consumers who evaluate different plans\cite{Fresno-ArandaFD22}.
    \item Limitation-Aware Testing: testing plays a pivotal role in ensuring the reliability and performance of an API. However, testing needs to be done keeping in mind the limitations set for each API tier. For example, one might need to simulate a real-world scenario where multiple mock consumers try to access the API simultaneously, to verify if the set limitations are truly upheld, ensuring fair service distribution among consumers.
    \item Automated Throttling Management: with the risk of overloading an API, there is a need for automated systems that can monitor and control the flow of requests to ensure compliance with set limitations. Throttling mechanisms, when efficiently managed, can act proactively to prevent any breaches of the contract, for example, by slowing down request acceptance once a certain threshold is approached.
    \item Configuration of API Gateways/Proxies: API gateways and proxies play an instrumental role in managing, monitoring, and securing API calls. They often need to be correctly configured to impose the right limitations based on the API consumer's tier or API key. Misconfigurations can lead to service breaches, unauthorized access, or even denial of service. Having a structured and standardized way to represent these limitations can significantly ease the gateway or proxy configuration process, ensuring adherence to the promised service levels.
\end{itemize}


In order to deal with this gap, the objectives of the present study are the following:
(i) to model with rigour the concept of limitation in the context of a RESTful Web APIs, and to study its validity properties;
(ii) to provide a specific serialization of the model aligned with the current \textit{de facto} OpenAPI Specification (OAS) standard to boost the ecosystem of tools;
(iii) to analyze of the expressiveness of the model based on a systematic modeling of real APIs;
(iv) to define a validity operation for validating the description of API limitations; and
(v) to present a prototyping tool to automate the validation analysis.

The rest of this paper is structured as follows: Section~\ref{sec:pricingAPIEconomy} presents the motivation behind our proposal by discussing a real scenario, and introduces the vocabulary used in the industry; Section~\ref{sec:pricingModel} sets out a pricing model and a corresponding OAS-aligned serialization; Section~\ref{sec:analysis} presents a validity operation; in Section~\ref{sec:validation} we validate the expressiveness of our model by reviewing 268 APIs in the industry to define a representative subset of 54 APIs that is modeled and analysed with our model, and we describe the tools developed to automate the analysis operation;  Section~\ref{sec:relatedWork} describes related work; and Section~\ref{sec:futureWork} and Section~\ref{sec:conclusions} present some conclusions and final remarks.

\section{Pricing in the API Economy}
\label{sec:pricingAPIEconomy}

In the API Economy world, API providers have to make
sufficient information available for the consumer to get informed about
their products. This includes information regarding the API
itself (endpoints and methods), the \textit{plans} that a
user can subscribe to, and the associated \textit{cost}. A
\textit{plan} includes information regarding the API's
limitations (\textit{quotas} and \textit{rates}) for each
of its resources.

All this information is typically found in a section called
\textit{pricing}; consequently, we shall henceforth consider 
a \textit{pricing} to be a set of \textit{plans} having an 
associated \textit{cost}.

\begin{figure}[hbt!]
    \centering
    \includegraphics[width=1\linewidth]{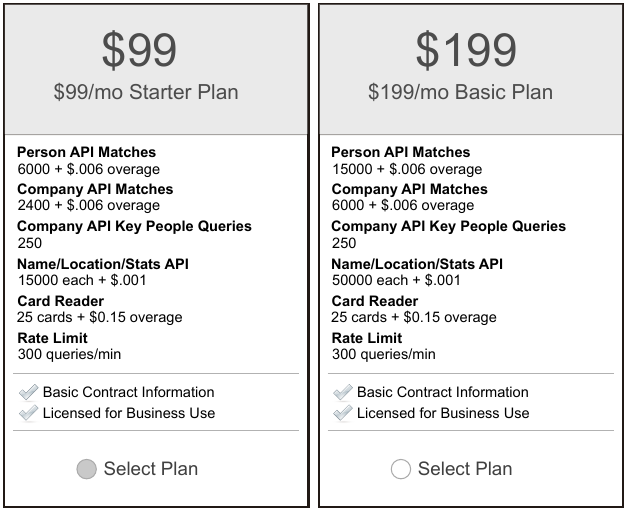}
    \caption{Plans of the FullContact API.}
    \label{figure:realPricingPlan}
\end{figure}

In order to illustrate these concepts, we present a real
example: the \textit{FullContact} API -- a tool for managing and
combining contacts from different sources (Gmail, social
media, etc.). The API allows users to programmatically look
up information and to match email addresses with publicly
available information so as to enrich the contacts. 
Figure~\ref{figure:realPricingPlan} depicts the pricing
extracted from the
\textit{FullContact} \cite{fullcontactarchive} API.

This \textit{pricing} example consists of two paid
\textit{plans} having a fixed \textit{cost} billed
monthly.  With respect to the \textit{limitations}, for each
\textit{operation}, a \textit{quota} is applied. For
example, in the Starter \textit{plan}, only 6000 matches on Person are available. Nevertheless, an \textit{overage}
is defined, i.e., it is possible to surpass the
\textit{limit} by paying a certain amount of money, in this
case, \$0.006 per request. Regardless of the plan,
a common \textit{rate} of 300 queries per minute is
applied.

In this context, several analytical challenges can arise
since the API providers need to understand 
the plans in depth before taking further action. In particular, they should verify the validity of
their plans (i.e., that there is nothing
inconsistent).

Those challenges correspond to common questions on the
API's pricing and plans that could be answered automatically
with an appropriate model and analytical framework providing
different analysis operations. The following two sections will detail
 the proposed model (Section~\ref{sec:pricingModel}) and the definition of a validity operation (Section~\ref{sec:analysis}).


\section{Pricing Model}
\label{sec:pricingModel}

In this section, we first present the Pricing4APIs model (Section~\ref{sec:governify4APIsModel}), then introduce SLA4OAI (Section~\ref{sec:SLA4OAI}), a specific textual serialization compatible with the OpenAPI Specification.
Both sections will use the FullContact pricing described above as a running example.

\subsection{The \textit{Pricing4APIs} Model}
\label{sec:governify4APIsModel}

As a high-level overview, the \textit{Pricing4APIs} model allows to define a set of plans with its associated cost; for each plan, a set of limitations (i.e. quotas and
rates) over the potential API operations can be defined. In the context of the RESTful paradigm, those operations are bounded to an HTTP path
and method.

\begin{figure*}[thb!]
    \centering
    \includegraphics[width=0.75\linewidth,trim={1cm 7cm 7cm 1cm},clip]{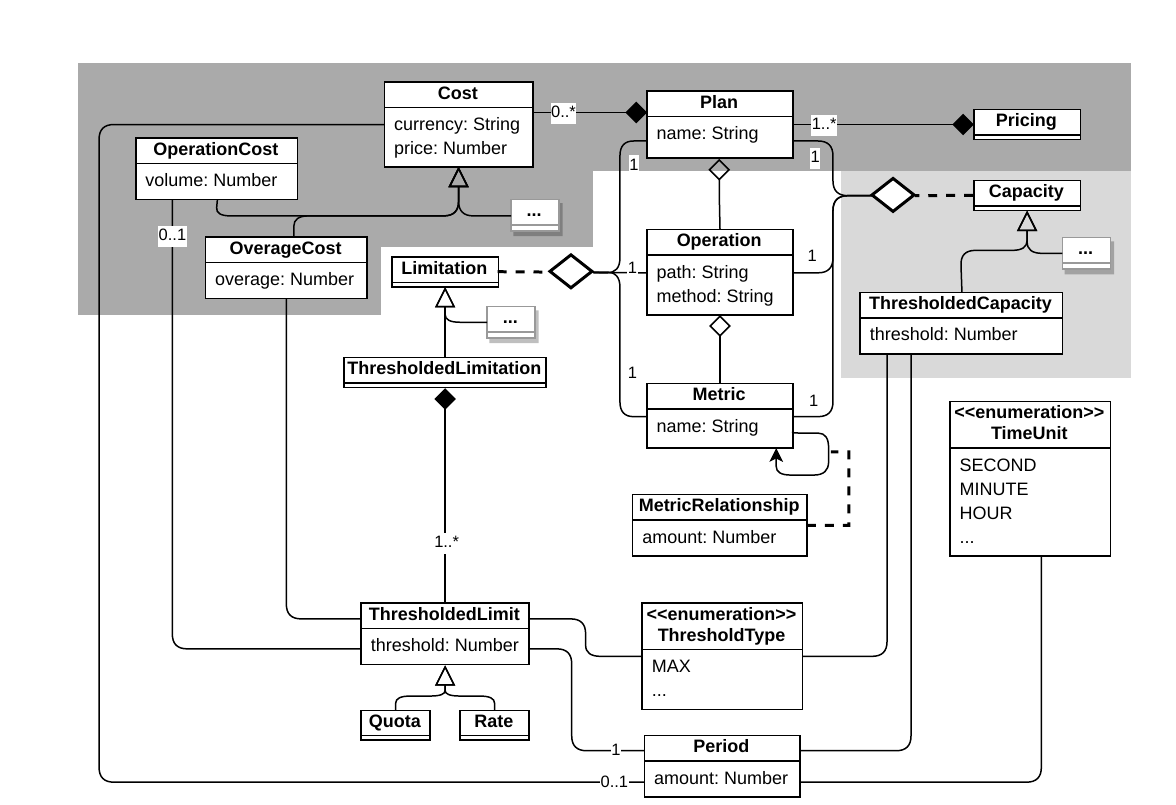}
    \caption{\textit{Pricing4APIs} model for API pricing.}
    \label{fig:PricingLimitationModel_OK_1_gray}
\end{figure*}

Figure~\ref{fig:PricingLimitationModel_OK_1_gray} depicts
the entire \textit{Pricing4APIs} model. For the sake of
clarity, we have split it into three areas: (i)
the dark gray area, \textit{pricing, plans and cost}; (ii) the unshaded area, \textit{limitations and limits}; and (iii) and the light gray area, \textit{capacity}.
In the following subsections, we will detail each part of the model with examples extracted from the FullContact API in Figure~\ref{figure:realPricingPlan}, considering each part: the plan area (Subsection~\ref{sec:pricingPlansCost}), the limitations area (Subsection~\ref{sec:limitationsLimits}), and the capacity area (Subsection~\ref{sec:capacity}).



\subsubsection{Pricing, Plans, and Cost}
\label{sec:pricingPlansCost}



As depicted in the model (dark grey area in Figure~\ref{fig:PricingLimitationModel_OK_1_gray}), a \texttt{Pricing} consists of a set of \texttt{Plans}. A \texttt{Plan} has a name and a \texttt{Cost} that defines the price charged to users so that they can access the service. In our example, the FullContact API has two \texttt{plans}: a \textit{Starter} and a \textit{Basic} \texttt{Plan}.

The \texttt{Cost} may be very simple (e.g., assign a constant price to the \texttt{Plan}, such as \textit{\$99} or \textit{\$199} in our example) or may depend on other properties. 
In this latter case, when the cost depends on a \texttt{Limitation}, we distinguish two costs:  \texttt{OperationCost}, when an \texttt{Operation} is being charged for each time it is invoked; and \texttt{OverageCost}, when once a certain value of the \texttt{Limitation} has been reached (cf. Subsection~\ref{sec:limitationsLimits}), there start to be imposed charges per volume.

Either type of \texttt{Cost} can be periodic, defining a \texttt{Period} with an amount and a \texttt{TimeUnit}. In our example, the \texttt{Cost} of the \textit{Starter} \texttt{Plan} is \textit{\$99} billed \textit{monthly}, i.e., it has a \textit{Period} with value 1 of the \textit{TimeUnit} MONTH.

An \texttt{OperationCost} is frequent in pay-as-you-go payment models in which there is no monthly fixed \texttt{Cost} and the API consumer is only charged for, given a \textit{requests} \texttt{Metric}, the number of requests. In the model, this cost is associated with the operation by means of the \texttt{Limitation}. For example, a service might offer a \texttt{Plan} A in which each request can be charged at \$0.10 (volume: 1) and a \texttt{Plan} B where each pack of 1000 requests (volume: 1000) is charged at \$75. Depending on the client's needs, they might prefer \texttt{Plan} A or \texttt{Plan} B.

An \texttt{OverageCost} is usual when providers do not want to cut off the service once a \texttt{Limitation} has been reached but want to continue providing it at a certain charge. Our example defines an overage when the \texttt{quota} values are reached: \textit{each additional match after 6000 monthly matches is charged at \$0.006}.





\subsubsection{Limitations and Limits}
\label{sec:limitationsLimits}



As depicted in the model (unshaded area in
Figure~\ref{fig:PricingLimitationModel_OK_1_gray}), in order
to carry out this regulation of the consumption of an API,
each \texttt{Operation} in a \texttt{Plan}  can be subject
to \texttt{Limitations} on a \texttt{Metric}. The most frequent type of \texttt{Limitation} 
is the \texttt{ThresholdedLimitation} which establishes one or
more \texttt{ThresholdLimits} on the number of
\texttt{Metric} units in a \texttt{Period}. The
\texttt{ThresholdType} is usually MAX (i.e., the
\texttt{ThresholdLimit} would therefore represent the
\textit{maximum} number of \texttt{Metric} units allowed).  In defining their \texttt{Pricing},
\texttt{Limitations} allow
providers to
adjust the  API's consumption to  the platform's total \texttt{Capacity} (cf.
Subsection~\ref{sec:capacity}).

An \texttt{Operation} is defined by the pair formed by HTTP method
and path. For example, \textit{GET /contacts} would
represent the query operation on a collection of user-type
objects.  A common example of a \texttt{Metric} is the
\textit{number of requests}. Nevertheless, other metrics
can be defined such as \textit{storage}, \textit{bandwidth} or \textit{CPU consumption}. Two \texttt{Metric}s can have a relationship that can be explicitly set in order to be taken into account in the plan validation; as an example, we can think in a escenario where each request to the API consumes 2KB of bandwith, consequently the \texttt{MetricRelationship} between the \texttt{Metric}s \textit{number of requests} (adimensional) and \textit{bandwith per request} (measured in KB) would be \textit{2}.

Depending on the algorithm used to update the \texttt{Limitations} \texttt{Metric}, we identify two types of \texttt{ThresholdedLimit}: on the one hand, the 
 \texttt{ThresholdedLimit} is classified as \texttt{Quota} if the
computation of the number of metric units is done over a
static window, i.e., in a \textit{fixed} time window.  For example,
a \textit{one-week static window} might be such that it always starts on
Monday at 00:00 and ends on Sunday at 23:59, regardless of
when the first metric unit is computed.
On the other hand, if the time window is \textit{sliding},
i.e., relative to the first metric unit computed, the \texttt{ThresholdedLimit} is classified as \texttt{Rate}. 
For example, in a \textit{one-week sliding window}, if the
first metric unit were computed on Wednesday at 15:36:39,
that window would close on the following Wednesday at
15:36:38.

\begin{figure}[hbt!]
    \centering
    \includegraphics[width=1\linewidth]{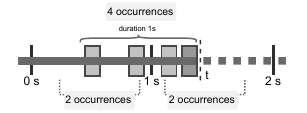}
    \caption{Sliding (rates) vs static (quotas) windows.}
    \label{fig:quotaVSRate}
\end{figure}

Figure~\ref{fig:quotaVSRate} illustrates graphically the
differences between sliding and static windows in a hypothetical scenario for the \texttt{Metric} of \textit{number of requests} (i.e., each gray box in the figure represents a request over the API \texttt{Operation} involved in the \texttt{Limitation}).
Considering the instant \textit{t} when the last request
was made, the analysis of the situation is twofold: (i)
inspecting 1 second back, i.e., a 1-second sliding window,
there exist 4 occurrences; (ii) observing only the 1-second
static window elapsed from 0s to 1s and from 1s to t, there
only exist two occurrences.  In short, depending on whether
a sliding (rate) or a static (quota) window is chosen, the observed
occurrences may differ.

In such an example, if we want to prevent our users from making more than 4 requests per second, there are two different alternatives: a 1-second sliding time window with a limit of 4 requests, that opens after the first request and prevents more than 4 from being made during that second; or a 1-second static window with a limit of 2 requests, that could concentrate the first two requests at the end of the first second and the other two at the beginning of the next one.

In the industry, the usage of this types of \texttt{ThresholdedLimit}, tend to follow patterns~\cite{Gamez-Diaz2017_icsoc_main}. Specifically,
\texttt{Quotas} tend to be defined over all sort of \texttt{Metric} and are
measured in periods longer than an hour (e.g., daily, weekly, monthly or yearly),
while \texttt{Rates} tend to be defined
over the specific \texttt{Metric} of \textit{number of requests} and are measured in shorter periods (e.g., secondly or minutely). Unfortunately, in most cases, the type of \texttt{ThresholdLimit} is not explicitly defined in the documentation and API consumers should determine the type by manually testing the API.  

In our FullContact example, the \textit{Starter}
plan has one \texttt{Rate} (\textit{300 requests in a
1-minute sliding window}) and four different \texttt{Quotas} (e.g. \textit{6000 matches in a 1-month static window}).

The model distinguishes two concepts:
\texttt{ThresholdedLimitation} and
\texttt{ThresholdedLimit}. A \texttt{ThresholdedLimitation}
over a certain metric and operation establishes a fraction
of the overall \texttt{Capacity} of the service. 
A \texttt{ThresholdedLimitation}, however, can be expressed in various
ways, one of which is by defining a set of
\texttt{ThresholdedLimits} that, within a time
period,  restrict the percentage of \texttt{Capacity} that consumers
are allowed to use.  For example, a
\texttt{ThresholdedLimitation} on a certain operation can be
defined as a set of \texttt{ThresholdedLimits} as follows:
\textit{30 requests every 1 week} and \textit{1 request
  every 1 second}.
We can find other alternatives to a set of \texttt{ThresholdedLimits} to express a \texttt{Limitation} and, consequently, we leave an appropriate extension point in the model (represented by the squared ellipsis "..."); for instance, we could express a \texttt{Limitation} using frequency distributions~\cite{SREBook}: in this way, 
percentiles could be used to define the form of the distribution and its different attributes. In such an example, a percentile such as 99.0 or 99.9 would represent a plausible value in the worst case, while the 50th percentile would emphasize the typical case. In this paper, however, we will focus only on \texttt{ThresholdedLimits} as they are the most prominent ones found in the industry.


\subsubsection{Capacity}
\label{sec:capacity}


Finally,  a crucial aspect that is not explicitly
depicted in a pricing or a plan is the \texttt{Capacity}. This
is an internal aspect that providers do not put out publicly.
The \texttt{Capacity} of the service represents a subset of the
constraints of the platform or system on which the service is
being deployed. It is the result of having to satisfy mainly technical and budget
criteria (e.g., CPU or memory, number of nodes of the
cluster, etc.).

Estimating the service's \texttt{Capacity}  is
fundamental to defining the \texttt{Pricing} and analyzing the
\texttt{Limitations}. In particular, all the \texttt{Limitations}
ought to be satisfied by the service, i.e., they must not
exceed the service's \texttt{Capacity}.


As depicted in the model (light grey in
Figure~\ref{fig:PricingLimitationModel_OK_1_gray}), once the
\texttt{Capacity} has been identified, it is specified
as if it were a \texttt{Limitation}, i.e., the number of
certain \texttt{Metric} units in a given \texttt{Period}.
Therefore, analogously to the \texttt{Limitation}, the
\texttt{ThresholdedCapacity} has a threshold value and a
\texttt{ThresholdType} (usually MAX) in a given
\texttt{Period} of a \texttt{TimeUnit}.

A possible way to express the \texttt{Capacity} on the metric \textit{request} is the
\textit{number of requests per second (RPS)} for each
operation and plan. For example, a capacity of
\textit{10\thinspace000 RPS} in \textit{GET /contacts} in the
\textit{free plan} would mean that the entire set of
free-plan users will be able to make 10\thinspace000 RPS.  The \texttt{Capacity} can be different for each
plan since different infrastructures may be used to provide a better level of  service to the
clients.

For example, an organization might have calculated, based on
performance and stress tests, that its production cluster is
able to accept 10\thinspace000 RPS. Consequently, if a
limitation had been set  of 10 requests per second per
client, the theoretical number of concurrent requests would
be $10\thinspace000/10=1000$ concurrent clients.

A useful instrument when analyzing \texttt{Limitations}  is
the \textit{percentage of capacity utilization} or simply
the \textit{percentage of utilization (PU)}.  Intuitively,
this percentage directly determines whether or not a
\texttt{Limitation} can be set  because this will be impossible if the PU is
greater than 100\%.

The PU will depend on how a consumer consumes the API. There
are two interpretations given a \textit{Limitation}: uniform
and burst. Therefore, the PU can be calculated in two
different ways.  To illustrate this idea, let us consider
a \texttt{ThresholdedLimitation} with a single
\texttt{ThresholdedLimit} of \textit{43\thinspace200
  requests every 1 day}:

In a first approximation, an API consumer could assume that,
since 1 day is 86\thinspace400 seconds, for every second,
they will have $43\thinspace200/86\thinspace400=0.5$
requests. In this case, it is assumed a \textit{uniform
  distribution} in which, little by little, the consumer
will reach the 43\thinspace200 requests available in the
day. This scenario corresponds to the \textit{minimum PU}.
But the \texttt{ThresholdedLimitation} states that for 1 day
it is possible to make 43\thinspace200 requests, and in no
case does it prevent the consumer from making all of them
in a burst in the first instant of time. Indeed, in 1 second
the consumer could make the whole set of 43\thinspace200
requests. This scenario implies a \textit{burst
  distribution}, and corresponds to the
\textit{maximum PU}.

Consequently, the PU must take both these models into
account, so that we define the \textit{bounded PU (BPU)}
as this range:

\begin{enumerate}
    \item The lower bound is the \textit{minimum PU}, in which a \textit{uniform} distribution of utilization over the period is assumed.
    \item The upper bound is the \textit{maximum PU}, which assumes the utilization of the maximum allowed in a single \textit{burst}.
\end{enumerate}

\begin{figure}[hbt!]
    \centering
    \includegraphics[width=1\linewidth]{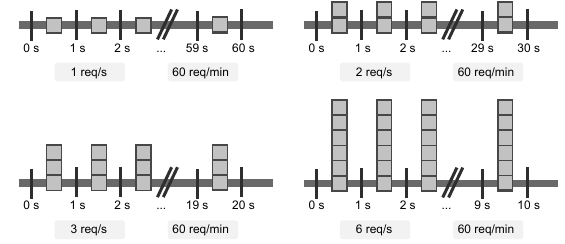}
    \caption{Examples of different consumption scenarios for the same \texttt{ThresholdedLimitation}.}
    \label{fig:consumption}
\end{figure}

Figure~\ref{fig:consumption} illustrates different consumption scenarios for the same \texttt{ThresholdedLimitation} of \textit{60 requests every 60 seconds}.

In a \textit{uniform} consumption, 60 requests in 60 seconds would be equivalent to 1 request every 1 second.
However, in a \textit{burst} consumption, 2, 3, 6, or even a maximum of 60 requests could be made in 1 second.
Therefore, to calculate the BPU in the limitation of \textit{60 requests every 60 seconds}, we should take as a minimum value the \textit{uniform} distribution of 1 request per second and as a maximum value the \textit{burst} of 60 requests in 1 second in a 1 minute window.

In this point, it is important to highlight that accurate capacity analysis stands as a cornerstone for ensuring service reliability and optimal resource utilization in the ever-evolving landscape of the API Economy. In environments where providers are catering to a diverse clientele, each contracting different plans, and thus varying service levels, the challenges multiply. For instance, consider multiple consumers, each having different allowances in terms of requests per second. The provider must not only ensure that each consumer gets their entitled service level, but also needs to manage back-end resources efficiently to serve everyone without degradation in service quality. Such intricate balancing acts require a deep understanding of capacity. Capacity misestimations can lead to overprovisioning, resulting in wasted resources and increased costs, or underprovisioning, leading to service disruptions and potential revenue loss. Moreover, in cloud-native environments, where elasticity is a prized feature, driving elasticity rules based on accurate capacity analysis becomes paramount. By knowing the exact capacity and workload distribution across different service tiers, elasticity rules can be defined to dynamically scale resources up or down. This not only ensures a consistent quality of service, regardless of the workload fluctuations of various plan subscribers, but also leads to more efficient and cost-effective operations. In essence, accurate capacity analysis underpins the ability of providers to uphold their service commitments while operating in a resource-optimized manner, making it a critical aspect of the API service paradigm.

\subsection{SLA4OAI: A Serialization for our Model}
\label{sec:SLA4OAI}


The \textit{Pricing4APIs} model can be serialized to be
aligned to a variety of API description specifications.
Specifically, we propose
SLA4OAI \cite{Gamez-Diaz2019_esecfse_industry,Gamez-Diaz2019_icsoc_main},
an extension of the OpenAPI Specification (OAS), as it is
currently the \textit{de facto} industrial standard for
describing APIs.  Nevertheless, our model could easily be
serialized to other API description languages (e.g., RAML,
API Blueprint, I/O Docs, WSDL or WADL).

It is important to highlight that, in the course of the last  years, we have led an interest group in the OAI Consortium, to recommend a first simplified version of SLA4OAI \footnote{\url{https://github.com/isa-group/SLA4OAI-Specification}} with the collaboration of 11 companies involving 22 practitioners. In this paper, we extend this simplified version to create an advanced specification \footnote{\url{https://github.com/isa-group/SLA4OAI-ResearchSpecification}} that incorporates a set of key novel features that allow the seralization of the full Pricing4APIs model: the globbing mechanism and extended costs models such as overage costs (present in 11.9\% of analyzed real-world pricings).

In SLA4OAI, the original OAS document is extended with an
optional attribute, \texttt{x-sla}, with a URI pointing to
the JSON or YAML document containing the SLA definition.
The SLA4OAI metamodel contains the following elements:
\textit{context information}, holding the main information
of the SLA context; \textit{infrastructure information}
providing details about the toolkit used for SLA storage,
calculation, governance, etc.; \textit{pricing information}
regarding the billing; and a definition of the \textit{metrics}
to be used. The main part of an SLA4OAI document is the
\textit{plans} section. This describes different service
levels, including the limitations set in the \textit{quotas}
and \textit{rates} sections. In what follows, we shall detail some of
the fields in a SLA4OAI file. Nevertheless, for a
comprehensive description of the syntax, a JSON Schema
document is available\cite{sla4oai-json-schema}. Further information is also available in the the specification's GitHub page.

As depicted in Listing~\ref{listing:sla4oaiHighLevel}, for
the SLA4OAI model, starting with the top-level element, one
can describe basic information about the \texttt{context},
the \texttt{infrastructure} endpoints that implement the
Basic SLA Management
Service~\cite{Gamez-Diaz2019_icsoc_main} (i.e., a protocol
as part of the SLA4OAI proposal, beyond the scope of the present paper), the \texttt{availability}, the \texttt{metrics} and,
inside \texttt{plans}, an entry defining \texttt{quotas},
\texttt{rates}, and \texttt{pricing}. Note that, in the model, the \texttt{pricing} of a \texttt{plan} is related to its cost and billing information.

\begin{lstlisting}[language=YAML,caption={Main elements in SLA4OAI},captionpos=b,label=listing:sla4oaiHighLevel]
  context: ...
  infrastructure: ...
  availability: ...
  metrics: ...
  plans:
    MyPlan:
      pricing: ...
      quotas: ...
      rates: ...\end{lstlisting}

Specifically, as depicted in Listing~\ref{listing:sla4oai_context}, the \texttt{context} contains general information, such as the \texttt{id}, the \texttt{version}, the URL pointing to the \texttt{api} OAS document, the \texttt{availability} of the document, and the \texttt{type} (this field can be either \texttt{plans} or \texttt{instance}). 
The \texttt{infrastructure} contains the endpoints that implement the Basic SLA Management Service, i.e., the \texttt{monitor} and \texttt{supervisor} services.

\begin{lstlisting}[language=YAML,caption={Context, infrastructure and availability details in SLA4OAI},captionpos=b,label=listing:sla4oai_context]
  context:
    id: FullContact
    sla: '1.0'
    type: plans
    api: ./fullcontact-oas.yaml
    provider: FullContact
  infrastructure:
    supervisor: https://...
    monitor: https://...
  availability: '2009-10-09T21:30:00.00Z'\end{lstlisting}
  
In the \texttt{metrics} field, as depicted in Listing~\ref{listing:sla4oai_metrics}, it is possible to define the metrics that will be used in the limitations, such as the number of requests or the bandwidth used per request. For each metric, the \texttt{type}, \texttt{format}, \texttt{unit}, \texttt{description}, and \texttt{resolution} (when the metric will be resolved, e.g., \texttt{check} or \texttt{consumption} to indicate that it will be sent before of after its consumption, respectively) can be defined.

\begin{lstlisting}[language=YAML,caption={Metric details in SLA4OAI},captionpos=b,label=listing:sla4oai_metrics]
  metrics:
  requests:
    type: integer
    format: int64
    description: Number of requests
    resolution: consumption
  matches:
    type: integer
    format: int64
    description: Number of matches\end{lstlisting}

The \texttt{plans} section, as depicted in Listing~\ref{listing:sla4oai_plans}, has the elements that will describe the plan-specific values -- \texttt{quotas}, \texttt{rates}, and \texttt{pricing}.

In this context, it is important to stress that the \texttt{plans} section maps the structure in the OAS document so as to attach the specific limitations (quotas or rates) for each path and method. In particular, the limitations are described with a \texttt{max} value that can be accepted and a \texttt{period} with \texttt{amount} and a time \texttt{unit}. 
Furthermore, the \texttt{cost} section defines the \texttt{overage} (including the \texttt{overage} threshold and \texttt{cost} per extra unit) and the \texttt{operation} (including the \texttt{volume} and the \texttt{cost} per unit) costs.

The SLA4OAI model supports globbing to simplify pricings where the same limitation applies to multiple paths and/or methods. The character \texttt{*} can be used as a wildcard, so that, for example, limitations attached to \texttt{'/v3/*'} apply to all paths starting with \texttt{'/v3/...'}, but not to \texttt{/api/v3/...}. For methods, limitations attached to method \texttt{all} will apply to all methods. It is worth noting that more restrictive globbed paths have higher priority than less restrictive paths, meaning that if they have limitations for the same metrics and methods, the limitations in the former will override the limitations in the latter. For example, \texttt{'/v3/operation/*'} has higher priority than \texttt{'/v3/*'}.

\begin{lstlisting}[language=YAML,caption={Plans details in SLA4OAI},captionpos=b,label=listing:sla4oai_plans]
  plans:
    Starter:
      pricing:
        cost: 99
        currency: USD
        period:
          amount: 1
          unit: month
      quotas:
        'v3/person.enrich':
          post:
            matches:
              - max: 6000
                cost:
                  overage:
                    overage: 1
                    cost: 0.006
      rates:
        'v3/person.enrich':
          post:
            requests:
                - max: 10
                  period:
                    amount: 1
                    unit: month\end{lstlisting}

\section{Analysis}
\label{sec:analysis}

In this section, we propose an analysis framework to form a ground on which to reason about the pricing
model presented.  Consequently, this framework paves the way
to exploiting the information contained in the model, and
 has been used to develop a validity analysis
operation that could be useful in a real setting for both consumers
and providers of APIs. As a foundation for the analysis
operations, the first of the following subsections addresses the
cornerstone of the analysis framework -- the relationship
between limitations and capacity. The subsequent subsection
will detail and exemplify the validity
operation that has been defined.

\subsection{Limits as Percentages of Capacity Utilization}

Since the capacity of the platform on which the service is
 deployed is not unlimited, the pricings should be defined to be compatible
and coherent with that capacity.  As an example, ensuring
that the total capacity is sufficient for the potential use
of the service defined in a particular plan should be
analysed. Furthermore, we proposed  (in Subsection
\ref{sec:capacity}) the notion that any given
limitation corresponds to a Bounded set of Percentage of
capacity Utilization (BPU) values derived from the potential usage
scenarios a client could have for their consumption within the API while
meeting its limitation.

In this context, the correspondence between limitations and
BPU can be obtained by means of a normalization procedure
that transforms the unit of the limitation to the capacity
time unit, and then computing the minimum and maximum possible PUs.
This procedure comprises just simple calculations, as is illustrated in the following
example:




Consider a
  limitation with a limit of \textit{43\thinspace200
    requests / 1 day} and assume a total capacity of
  50\thinspace000 RPS. Since all limitations should be
  expressed using the time unit of the capacity (second),
  the limitation is \textit{43\thinspace200 requests /
    86\thinspace400 seconds}. First, assume a
  \textit{uniform} consumption, i.e., if in 1 day (86\thinspace400
  seconds) there are 43\thinspace200 requests,
  there will be $43\thinspace200/86\thinspace400=0.5$
  RPS. Given the value of the capacity,
  50\thinspace000 RPS, the minimum PU is
  $0.5/50\thinspace000=0.000\thinspace01=0.001\%$. Now
  assume a single \textit{burst} consumption, i.e., if a burst of
  43\thinspace200 requests can occur during any 1 second window
  over 1 day. Given the value of the capacity, 50\thinspace000 RPS,
  the maximum PU is
  $43\thinspace200/50\thinspace000=0.864=86.4\%$. Therefore,
  the BPU of \textit{43\thinspace200 requests / 1 day} subject
  to a capacity of 50\thinspace000 RPS is [0.001\%-86.4\%].

The calculation of the overall system capacity is a
nontrivial procedure. It requires great technical effort to make a proper estimate.  But, depending on the
stage of development, even this will not always be feasible.
In the present study, when the value of the system's capacity is unknown, we
shall define a default capacity as the value of the highest capacity needed for the case of a single consumer with a maximum PU. This assumption is a clear simplification that allows for a validation analysis with the uncertainty over the real capacity; however, in real scenarios, API providers are expected to have a much bigger capacity that supports multiple consumers of different plans simultaneously. To
calculate the default capacity value, we shall assume a uniform consumption
after normalizing to the smallest time unit, and take the
greatest value. As an example, let us assume the following two limitations:
  \textit{1 RPS} and \textit{100 RPW} (1 week, 604\thinspace800 s).
  In order to take the value
  of the highest capacity needed, we must first determine what
  the strongest limitation is. For this case, we normalize to the
  smallest unit, the second, $1\:RPS=1$ and
  $100\:req/604\thinspace800\:s=0.000\thinspace165\:RPS$,
  since $1>0.000\thinspace165$ we have that the
  highest capacity needed is \textit{1 RPS}.
  Therefore, we will take
  \textit{1 RPS} as the value of the capacity.


\subsection{Pricing Validity}

This section will detail a validity framework of the pricing plans, independent of the number of API consumers. It is important to note that analyses dependent on the number of users (i.e., validity of the plans and the capacity with a particular scenario of API consumers) are left for future work. The primary objective here is to determine whether there are consistencies in the pricing plans or not. For the sake of clarity, the examples provided in this section are deliberately simplified. However, it should be emphasized that, in real-world pricing scenarios, the process of checking validity can become quite complex.

Specifically, we define the \textbf{validity} of a pricing as checking whether it is valid depending on a set of validity criteria that represent the absence of different types of conflict, for example, \textit{two limits within a limitation that cannot be met at the same time}.

Consequently, the validity of a \textit{Pricing4APIs} model is defined
as certain validity criteria being met in each part of the
model. In the model, a \textit{pricing} has a set of
\textit{plans}, and these plans consist of
\textit{limitations}, each with its own
\textit{limits}. This hierarchy carries over to the
validity operation.  Hence, for example, a \textit{pricing}
will be valid, notwithstanding its satisfying other additional validity
criteria, if all of its \textit{plans} are also valid.

For solving validity conflicts, a \textbf{priority criterion} is required. For example, \textit{if two limits are defined with different values for a given metric and operation, which one should prevail over the other?} In order to satisfy these requirements we assume henceforth the following default priority criteria: i) limitations with smaller periods over limitations with higher periods; ii) rates over quotas; iii) metric \textit{number of requests} over any other metric. Nevertheless, these criteria can be re-defined in other scenarios (e.g., metric \textit{requests} may be less important than the bandwidth in a certain business context).

\begin{figure*}[tb!]
    \centering
    \includegraphics[width=1\linewidth]{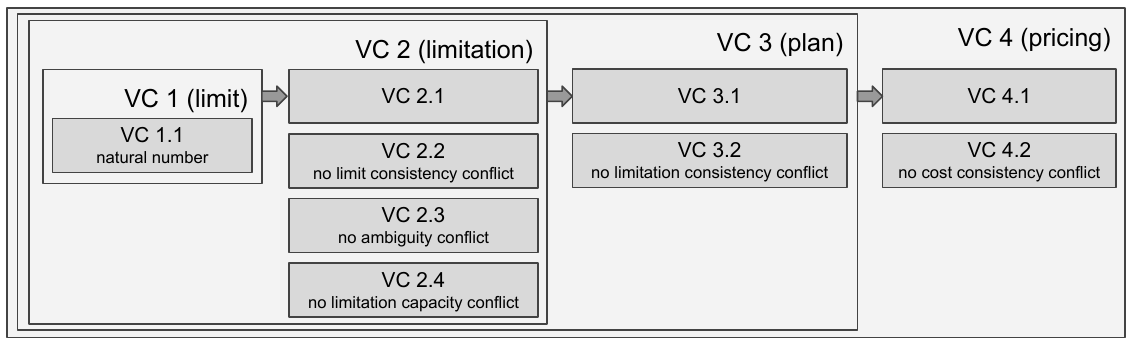}
    \caption{Validity criteria hierarchy.}
    \label{fig:ValidityHierarchy}
\end{figure*}

We shall present the validity criteria in a hierarchy,
starting from the fine-grained (VC1 - limits, VC2 -
limitation) to the coarse-grained (VC3 - plan, VC4 -
pricing)  validity criteria. Each validity criterion comprises
multiple validity subcriteria. 
Figure~\ref{fig:ValidityHierarchy} gives an overview of
this hierarchy of validity criteria.
The details of each validity criterion are as follows:

\begin{description}
    \item[VC1 - Valid limit] A \textit{limit} is valid if
              its \textit{threshold} is a natural number (VC1.1).
    \item[VC2 - Valid limitation] A \textit{limitation} is valid if:
              all its \textit{limits} are valid (VC2.1);
              there are no \textit{limit consistency conflicts} between any pair of its \textit{limits}, i.e., there is no situation exceeding a \textit{limit} with less priority while it is allowed by another \textit{limit} with higher priority (VC2.2);
              there are no \textit{ambiguity conflicts} between any pair of its \textit{limits}, i.e., two limits using the same period with different values (VC2.3) and
              there is no \textit{capacity conflict}, i.e., the limitation does not surpass the associated \textit{capacity} (VC2.4).
    \item[VC3 - Valid plan] A \textit{plan} is valid if:
              all its \textit{limitations} are valid (VC3.1) and
              there are no \textit{limitation consistency conflicts} between any pair of its \textit{limitations}, i.e., two \textit{limitations} on two related \textit{metrics} (by a certain factor) cannot be met at the same time (VC3.2). If they happen to exist, the priority criteria will be used for determining which limit has to be prioritized.
    \item[VC4 - Valid pricing] A \textit{pricing} is valid if:
              all its \textit{plans} are valid (VC4.1) and
              there are no \textit{cost consistency conflicts} between any pair of its plans, i.e., a \textit{limitation} in one plan is less restrictive than the equivalent in another \textit{plan} but the former \textit{plan} is cheaper than the latter (VC4.2).
\end{description}

In order to understand these validity criteria, on the following subsubsections we will present examples of existence and absence of conflicts of each type.

\subsubsection{Limit Consistency Conflict (VC2.2)} 

\begin{lstlisting}[
  language=YAML,
  caption=Validity criterion VC2.2 (no limit consistency conflict),
  label=lst:p1d1.1b,
  captionpos=b,frame=single,
]
  Capacity: 1000000 RPS
  Limitations:
    Quota: 100 requests / 1 day
    Quota: 1000 requests / 1 week                (VALID)\end{lstlisting}

An example of a situation where \textbf{there is no limit consistency conflict} can be observed in Listing~\ref{lst:p1d1.1b}. An inconsistency occurs when there is a possible situation exceeding a \textit{limit} with less priority while it is allowed by another \textit{limit} with higher priority, according to the priority criteria herein before mentioned.
An example, using the size of the periods as the priority criterion, a conflict shall happen if the maximum PU of the limit with the longest period is less than the maximum PU of the limit with the shortest period.

The limit having the longest period is \textit{1000 requests / 1 week} whose maximum PU is $1000/1\thinspace000\thinspace000=0.10\%$. The limit with the shortest period, \textit{100 requests / 1 day}, has a maximum PU of $100/1\thinspace000\thinspace000=0.01\%$. Since $0.10\% \not< 0.01\%$, there is no conflict between these limits.

\begin{lstlisting}[
  language=YAML,
  caption=Validity criterion VC2.2 (limit consistency conflict),
  label=lst:p1d1.1a,
  captionpos=b,frame=single,
]
  Capacity: 1000000 RPS
  Limitations:
    Quota: 100 requests / 1 day
    Quota: 10 requests / 1 week                (INVALID)\end{lstlisting}

On the other hand, in Listing~\ref{lst:p1d1.1a} \textbf{there is a limit consistency conflict}. 
The limit with the longest period is \textit{10 requests / 1 week} whose maximum PU is $10/1\thinspace000\thinspace000=0.001\%$. The limit with the shortest period, \textit{100 requests / 1 day}, has a maximum PU of $100/1\thinspace000\thinspace000=0.01\%$. Since $0.001\% < 0.01\%$, there is a limit consistency conflict between these limits.

\subsubsection{Ambiguity Conflict (VC2.3)} 

\begin{lstlisting}[
  language=YAML,
  caption=Validity criterion VC2.3 (no ambiguity conflict),
  label=lst:p1c2b,
  captionpos=b,frame=single,
]
  Limitation:
    Limit: 1 request / 1 second
    Limit: 100 requests / 1 day                  (VALID)\end{lstlisting}

An example where \textbf{there is no ambiguity conflict} is presented in Listing~\ref{lst:p1c2b}, because the limits of the limitation use different periods, i.e., \textit{1 second} and \textit{1 day}.

\begin{lstlisting}[
  language=YAML,
  caption=Validity criterion VC2.3 (ambiguity conflict),
  label=lst:p1c2a,
  captionpos=b,frame=single,
]
  Limitation:
    Limit: 1 requests / 1 second
    Limit: 100 requests / 1 second             (INVALID)\end{lstlisting}

Conversely, in Listing~\ref{lst:p1c2a} \textbf{there is a consistency conflict} because the limits of the limitation use the same period, i.e., \textit{1 second}.

\subsubsection{Capacity Conflict (VC2.4)}
\begin{lstlisting}[
  language=YAML,
  caption=Validity criterion VC2.4 (no capacity conflict),
  label=lst:p1c1.2b,
  captionpos=b,frame=single,
]
  Capacity: 100 requests / 1 second (100 RPS)
    Limitations:
      Quota: 50 requests / 1 day                 (VALID)\end{lstlisting}

A possible situation where \textbf{there is no capacity conflict} is shown in Listing~\ref{lst:p1c1.2b}.
First, we normalize using the unit of the capacity (i.e., seconds). Thus, there are \textit{50 requests / 86\thinspace400s (1 day)}.
Next, to calculate the BPU, we need both (i) the \textit{minimum PU} (uniform distribution) and (ii) the \textit{maximum PU} (burst distribution).
For (i), if in 86\thinspace400 seconds there are 5 requests, in 1 second there will be $50/86\thinspace400=0.000\thinspace57$ requests. The \textit{minimum PU} is $0.000\thinspace57/100=0.0057\%$.
For (ii), in 1 second there will be a burst of $50$ requests. The \textit{maximum PU} is $50/100=50\%$.
Therefore, the BPU is [0.0057\%,50\%].

Since BPU is always less than 100\%, there is no capacity conflict.

\begin{lstlisting}[
  language=YAML,
  caption=Validity criterion VC2.4 (capacity conflict),
  label=lst:p1c1.2a,
  captionpos=b,frame=single,
]
  Capacity: 100 requests / 1 second (100 RPS)
  Limitations:
    Quota: 200 requests / 1 day                (INVALID)\end{lstlisting}

On the contrary, in Listing~\ref{lst:p1c1.2a} \textbf{there is a capacity conflict}.
First, we normalize using the unit of the capacity (i.e., seconds). Thus, there are \textit{200 requests / 86\thinspace400s (1 day)}.
Next, to calculate the BPU, we need both (i) the \textit{minimum PU} (uniform distribution) and (ii) the \textit{maximum PU} (burst distribution):
\begin{itemize}
    \item For (i), if in 86\thinspace400 seconds there are 5 requests, in 1 second there will be $200/86\thinspace400=0.0023$ requests. The \textit{minimum PU} is $0.0023/100=0.0000\thinspace23\%$.
    \item For (ii), in 1 second there will be a burst of $200$ requests. The \textit{maximum PU} is $200/100=200\%$. Therefore, the BPU is [0.0000\thinspace23\%,200\%].
\end{itemize}
Since BPU is greater than 100\%, there is a capacity conflict because of the \textit{maximum PU}.

\begin{lstlisting}[
  language=YAML,
  caption=Validity criterion VC2.4 (no capacity conflict),
  label=lst:p1c1.2c,
  captionpos=b,frame=single,
]
  Capacity: 100 requests / 1 second (100 RPS)
  Limitations:
    Quota: 200 requests / 1 day
    Rate: 99 requests / 1 second                 (VALID)\end{lstlisting}

Additionally, Listing~\ref{lst:p1c1.2c} presents another example where \textbf{there is no capacity conflict}.
First, we normalize using the unit of the capacity (i.e., seconds). Thus, there are \textit{200 requests / 86\thinspace400s (1 day)} and  \textit{99 requests / 1s}
Next, we calculate the BPU in each limitation as in other examples.
The first limitation's BPU is [0.0000\thinspace23\%,200\%].
Next, to calculate the BPU, we need both (i) the \textit{minimum PU} (uniform distribution) and (ii) the \textit{maximum PU} (burst distribution).
For (i), the \textit{minimum PU} is $99/100=99\%$.
For (ii), in 1 second there will be a burst of $99$ requests. The \textit{maximum PU} is $99/100=99\%$.
Therefore, the BPU is [99\%,99\%].

Now, we aggregate both BPUs: first, we get the maximum value of the minimum PU: \textit{max(0.0000\thinspace23\%, 99\%)=99\%}.
Next, we obtain the minimum value of the maximum PU: \textit{min(200\%,99\%)=99\%}.

Therefore, as a result, we got [99\%,99\%]. Given that it does not surpass the capacity, we state that there is no capacity conflict.

\subsubsection{Limitation Consistency Conflict (VC3.2)} 

\begin{lstlisting}[
  language=YAML,
  caption=Validity criterion VC3.2 (no limitation consistency conflict by a related metric),
  label=lst:p1c3b,
  captionpos=b,frame=single,
]
  Limitation:
    Limit: 1000 KB / 1 month
  Limitation:
    Limit: 1000 requests / 1 month
  Relationship
    1 request = 0.5 KB                           (VALID)\end{lstlisting}

On the one hand, in Listing~\ref{lst:p1c3b} \textbf{there is no limitation consistency conflict by a related metric} because, if each request consumes 0.5 KB, in 1000 KB one would have at most $1000/0.5=2000$ requests. Given that $1000<2000$, the value of the limit on requests would not lead to any conflict.

\begin{lstlisting}[
  language=YAML,
  caption=Validity criterion VC3.2 (limitation consistency conflict by a related metric),
  label=lst:p1c3a,
  captionpos=b,frame=single,
]
 (Relationship: 1 request = 0.5 KB)
  Limitation:
    Limit: 1000 KB / 1 month
  Limitation:
    Limit: 5000 requests / 1 month             (INVALID)\end{lstlisting}

On the other hand, in Listing~\ref{lst:p1c3a} \textbf{there is a limitation consistency conflict by a related metric} because, if each request consumes 0.5 KB, in 1000 KB one would have at most $1000/0.5=2000$ requests. Since $5000>2000$, one could never reach 5000 requests, and there is therefore a conflict deriving from the relationship between metrics.


\subsubsection{Cost Consistency Conflict (VC4.2)} 

\begin{lstlisting}[
  language=YAML,
  caption=Validity criterion VC4.2 (no cost consistency conflict),
  label=lst:p1c4b,
  captionpos=b,frame=single,
]
  Plan 1:
    Limitation:
      Limit: 10 requests / 1 second
    Limitation:
      Limit: 100 requests / 1 day
    Cost: $10 / 1 month
  
  Plan 2:
    Limitation:
      Limit: 100 requests / 1 second
    Limitation:
      Limit: 1000 requests / 1 day
    Cost: $100 / 1 month                         (VALID)\end{lstlisting}

  An example where \textbf{there is no cost consistency conflict} can be observed in Listing~\ref{lst:p1c4b}, because any limitation in one of the plans
  is less restrictive than the equivalent in the other plan,
  but this other plan is also cheaper.  In this
  example,  plan 1 has stricter limitations and a lower
  cost than plan 2. The increase from 10 to 100 per-second
  requests, and from 100 to 1000 daily requests is also
  represented in the cost -- from \$10 to \$100.

\begin{lstlisting}[
  language=YAML,
  caption=Validity criterion VC4.2 (cost consistency conflict),
  label=lst:p1c4a,
  captionpos=b,frame=single,
]
  Plan 1:
    Limitation:
      Limit: 10 requests / 1 second
    Limitation:
      Limit: 100 requests / 1 day
    Cost: $10 / 1 month
  
  Plan 2:
    Limitation:
      Limit: 1 requests / 1 second
    Limitation:
      Limit: 1000 requests / 1 day
    Cost: $1 / 1 month                         (INVALID)
\end{lstlisting}

On the contrary, in Listing~\ref{lst:p1c4a} \textbf{there is a cost
  consistency conflict} in the two plans'
limitations and cost. While the decrease in
per-second requests  from 10 to 1 is indeed represented in the costs going from \$10
down to \$1, the increase from 100 to 1000 daily
requests is in the contrary direction to the decrease in costs. There is therefore a cost inconsistency.

\section{Evaluation}
\label{sec:validation}

In this section, we describe how we evaluated our
proposal to determine, on the one hand, the expressiveness of our model and
whether this is enough for it to express a wide variety of
real-world API pricings and to identify which characteristics of the real-world pricings are unable to be expressed; on the other hand, the automation potential of the validity analysis presented.  Specifically, we aim to answer the following two research questions (RQs):
\begin{itemize}
    \item \textbf{RQ1 - Expressiveness}. \textit{Is the modeling language expressive enough to model real-world API pricings?} We validated our language based on the analysis of two datasets containing a total of 268 selected APIs, with multiple pricings.
    \item \textbf{RQ2 - Automation}. \textit{Is it possible to automate the validation of API pricings?} Pricings should be valid and be devoid of inconsistencies between in their definition. We developed a tool to automate the analysis of API plans and solve the validity operation in \ref{sec:analysis}.
\end{itemize}


\subsection{RQ1 - Expressiveness}

\subsubsection{Analysing API Limitations and Pricing}
\label{sec:analyzingAPILimitationsPricing}

For this analysis, we considered three
different sources: (i) our work in Gamez-Diaz
et al.~\cite{Gamez-Diaz2017_icsoc_main} in which we analysed a
set of 69 APIs from two of the largest API directories; (ii)
the work of Neumann et al.~\cite{Neumann2018} in which the
authors analysed a set of 500 APIs from the top most popular 4000
websites in the Alexa ranking~\cite{alexa}; (iii) the 27 most popular APIs from RapidAPI\footnote{\url{https://rapidapi.com/collection/popular-apis}. Accessed on January 2023. Note that this list is regularly updated, and some APIs may be added, deprecated or removed. Some of the APIs included in our analysis are no longer available.}.

We adapted and applied the process described in
contribution \cite{Gamez-Diaz2017_icsoc_main}~(i), screening
the API repositories and applying the inclusion criterion
described by the authors (\textit{which includes more than 5000
  APIs with a last update in 2020}). The result
was the selection of one source: ProgrammableWeb. We
extracted the most popular API categories ($97th$
percentile, i.e., 14 categories selected, with more than
16\thinspace500 APIs). We filtered this dataset by
removing duplicates (only one API per company was
chosen at random). As a result, we had 2966 potential APIs to
study. Out of them, 30 APIs were selected.

In contribution \cite{Neumann2018}~(ii), the authors
analyzed a set of attributes of 500 APIs by focusing on
their general features such as their fit to REST best
practices and design decisions rather than on their specific
pricing aspects. Nevertheless, this dataset is interesting
as a starting point for our analysis since it includes a
variety of APIs and provides a comprehensive analysis of
certain attributes.  From this dataset, we filtered out
any rows which were not RESTful APIs, leaving a subset of
499 unique APIs.  First, we selected those APIs with a
\textit{Payment Plan}, as specified in a column in the
dataset, obtaining 55 APIs, which represents the 11.02\% of the total 499 APIs. We noted some
errata in the classification of some APIs that we are very
familiar with (e.g., GitHub was wrongly classified in the
"not having plans" section). The reason behind these errors might be that the APIs were analysed some time ago, when they did not have plans at the time; alternatively, some APIs might have their pricing plans \textit{hidden} within their documentation (e.g., we found that Yelp has an implicit VIP plan). This led us to analyze
the rest of the dataset (444) manually to check whether the
API still existed and whether it had API limitations. This
analysis resulted in 162 APIs to be included (67 with
pricing plans and 95 without but with API
limitations, which represent 15.09\% and 21.4\% respectively of these 444 APIs originally classified as "not having plans"). Adding these to the first set of 55 APIs, led to 217
APIs to be analyzed.

The list of most popular APIs from RapidAPI (iii) includes 27 different RESTful APIs. RapidAPI acts as a gateway to these APIs and provides simple pricing options. Some of these pricings differ from the pricings offered by the API providers in their official websites, which are often more complex. After a manual analysis of all of them, we found that 22 APIs had pricing plans and limitations, while 5 had no plans or limitations.

Combining the 30 APIs extracted according to~\cite{Gamez-Diaz2017_icsoc_main}, the 217 from the dataset in~\cite{Neumann2018} and the 22 from RapidAPI and removing duplicates left a dataset of 268 APIs -- the \textit{Pricing4APIs dataset}. Table~\ref{table:bigpic1} presents the overall  picture of the analysis that was carried out. Out of more than 17\thinspace027 APIs, we manually modeled 54 of them, having a 90\% confidence level and an 11\% margin of error \cite{Isserlis}. The full Pricing4APIs dataset, including details about their attributes, is available at \cite{modeled-dataset} as part of Dataset D01.

\begin{table}[!ht]
    \centering
    \caption{Main numbers of our Pricing4APIs dataset.}
    \label{table:bigpic1}
    \begin{tabular}{ll}
    APIs from Gamez-Diaz (N=2966)     & 30 \\
    APIs from Neumann (N=217)         & 217                         \\
    APIs from RapidAPI (N=27)         & 22
    \\
    Total APIs after removing dups.    & 268                                     \\
    APIs having pricing               & 176 out of  268                         \\
    Manually modeled APIs (N=266)     & 54 \\
    \end{tabular}
\end{table}


We analyzed 268 APIs in regard to two main types of attributes:
\textit{limitations} and \textit{pricing}. Both types
include a wide range of other attributes, some
supported by our model but others not. For example, our
model does support overage costs (e.g., \$0.1 per exceeded
request), but it does not support complex
metrics based upon HTTP protocol-related aspects (i.e.,
headers, parameters, etc.).

Although the Pricing4APIs dataset comprises 268 APIs,
only 176 of them, the 65.7\%, present a pricing or a plan. Consequently, the
analysis of pricing is limited to this reduced dataset.
Table~\ref{table:bigpic2} presents some results of the
analysis.

\begin{table}[!ht]
    \centering
    \caption{Results of the analysis in real-world APIs.}
    \label{table:bigpic2}
    \begin{tabular}{ll|l}
                                      & N=268                           & N=176 \\
    \hline
    Has limitations                   & 95.9\%                          & 94.9\%   \\
    Has quotas                        & 59.7\%                          & 72.2\% \\
    Has rates                         & 78.7\%                          & 69.9\% \\
    Has quotas and rates              & 42.5\%                          & 46.6\% \\
    Simple cost (e.g., monthly price) & 60.8\%                          & 92.6\% \\
    Has a pay-as-you-go cost model       & 9.3\%                        & 14.2\% \\
    Includes overage cost              & 11.9\%                         & 18.2\%
    \end{tabular}
\end{table}

\begin{description}
\item \textbf{Limitations analysis}: Most APIs (95.9\%) have
  limitations in terms of quotas (59.7\%) or rates (78.7\%).
  Almost half use a combination of the two (42.5\%).
  These limitations are usually rather simple (such as monthly
  requests for quotas and secondly requests for rates).
  However, a minority tend to have a higher level of
  expressiveness. For example, they use the information from
  the HTTP request -- from query parameters (2.2\%) to other
  low-level aspects of the HTTP message such as headers,
  body, etc. (9.3\%).  A marginal number of APIs allow
  consumers to exceed the limitation value once or many times
  per month (1.1\%).
\item \textbf{Pricing analysis}: the vast majority of the
  APIs  (92.6\%) include a simple (e.g., monthly) cost.
  Nonetheless, they may have operation costs (14.2\%) or
  include overage costs (18.2\%). Finally, a minority have
  purchasable add-ons or extras (5.7\%) or their pricing is
  calculated based on the number of users (10.2\%).
\end{description}

\begin{figure}[hbt!]
    \centering
    \includegraphics[width=1.\linewidth]{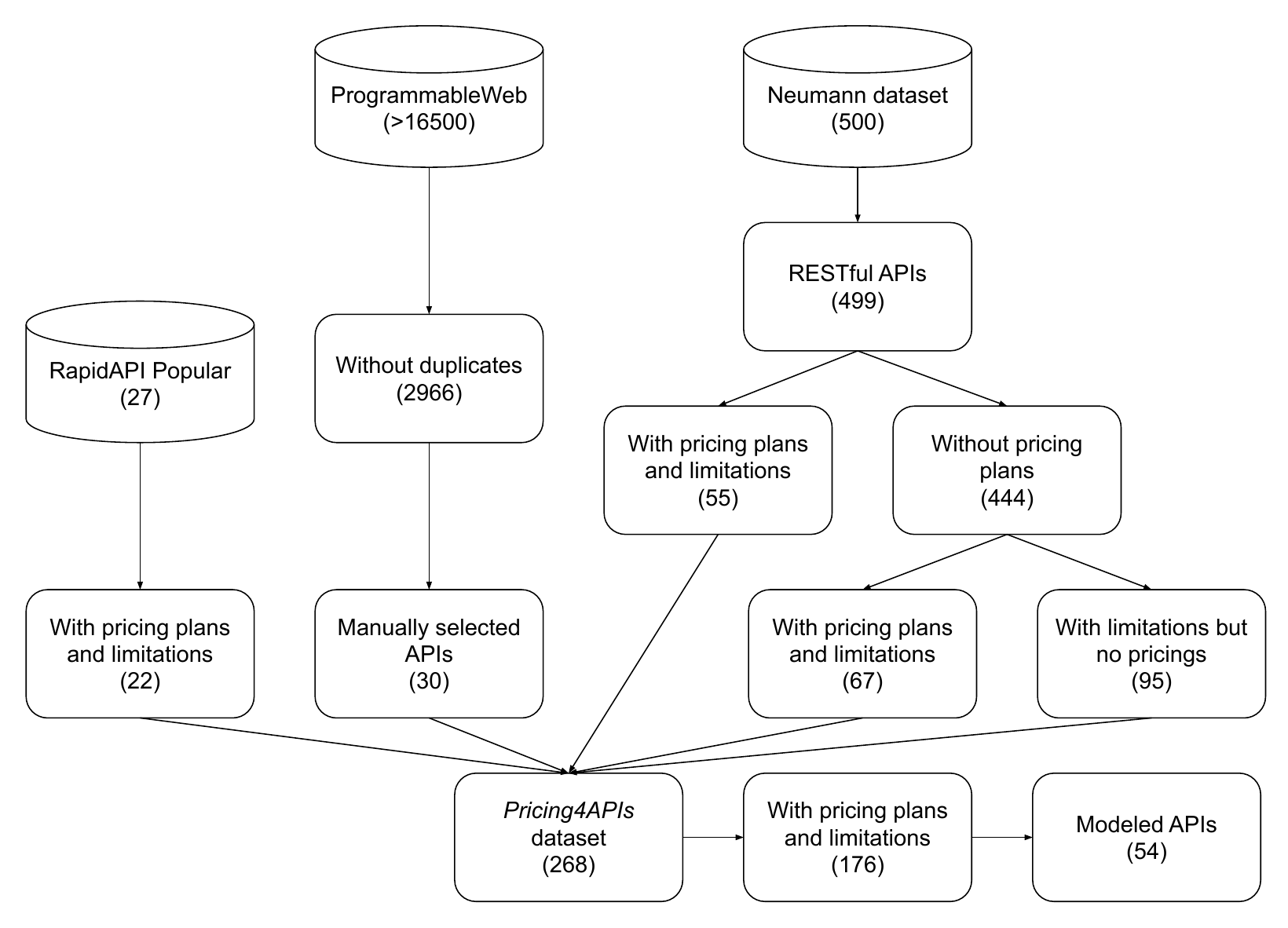}
    \caption{Diagram of the database filtering process, starting from the Gamez-Diaz and Neumann datasets and the RapidAPI most popular list.}
    \label{figure:diagdatabase}
\end{figure}


Given the 54 modeled APIs, we analyzed the different metrics included in the documentation of each of them. We found a total of 145 metrics, although different providers may name a same metric with different names (e.g. \textit{requests} and \textit{transactions}). 61.38\% of these metrics are domain independent (such as requests, storage or users), while 38,62\% are dependent (such as emails, documents or invoices). We grouped the 145 metrics in different categories based on their similarity, resulting in 14 categories. The most populated one is \textit{requests}, including 58 metrics. The second one is \textit{AI} (Artificial Intelligence), with 19 metrics, as some of the analyzed APIs are related to artificial intelligence and include a considerable amount of metrics. Many categories only include a few metrics because they are difficult to group together. This analysis is available at \cite{modeled-dataset} as part of Dataset D02.



\subsubsection{Modeling API Pricings}
\label{sec:modelingAPILimitationsPricing}

In this work, we seek to perform automated
operations through analyzers that solve the validity
operation regarding pricings, as were
discussed in Section ~\ref{sec:analysis}. Nevertheless,
before using any analyzer or tool, these
pricings have to be modeled.  This subsection
describes a validation of our Pricing4APIs model by
modeling a number of real-world APIs, first describing the
modeling process and
then the issues that arise during this process.  This process included
the construction of a curated list of 54 API
pricings with the model in subsection \ref{sec:SLA4OAI}, which represents the
variability found in the industry.


As noted above, we analyzed different attributes of the
Pricing4APIs dataset  of 268 APIs. The next step
would be to write the SLA4OAI specification of every API so that it can be passed to the
automated analyzer. However, since this is a time-consuming
task, we decided to follow a hybrid sampling approach, using purposive and probabilistic sampling \cite{baltes2022sampling}, to obtain a subset of APIs. With the former, we wanted to ensure that our model covers the most representative elements of API pricings, so we modeled all 22 APIs from RapidAPI's most popular list. With the latter, we aimed to reduce the threats of purposive sampling by modeling 32 additional pricings. According to \cite{baltes2022sampling}, purposive sampling is the most common approach in software engineering research, used in 76\% of studies. The resulting subset includes 54 APIs.

Note that the process of modeling a single API pricing
consists of (i) reading and understanding the entire API
documentation, (ii) extracting the API endpoints and methods
(skipped if OAS documentation is available), (iii) reading and
understanding every limitation in every plan of the API
pricing,  and (iv) specifying the metrics and API limitations
in accordance with our proposed model for each API path and
method. The process of modeling the API itself is tedious, which is
why APIs having a public OAS documentation greatly
facilitate the subsequent modeling task. With the introduction of globbing, step (ii) is simplified when the same limitation applies to multiple endpoints, and even completely unnecessary when the limitation applies to all endpoints. The 22 APIs from RapidAPI were modeled using globbing, which resulted in much simpler files.

In the following sections, we determine the issues found
during this OAS modeling
process.

In the process of modeling the pricings
of this subset of 54 APIs, we encountered several issues. We have
classified them into two categories: \textit{modeling
issues} and \textit{open issues}, depending on whether they are issues
that can be partially modeled with SLA4OAI or issues that need changes that will be taken into
consideration when establishing future work.

    \textbf{Modeling issues}: 
    
    \textit{MI-01} In \textit{pay-as-you-go} plans, users are
      only charged with the requests that they actually
      consume (e.g., \textit{FacePlusPlus}). This situation
      was modeled as a quota, with no \textit{max}
      field (or \textit{max: unlimited}) with its
      corresponding \textit{OverageCost}. As an alternative, we could also have modeled this as an \textit{OperationCost}. 
     
    \textit{MI-02} In some APIs (e.g.,
      \textit{FacePlusPlus}), the \textit{operation cost}
      depends on the HTTP status code that is returned to
      the consumer. Hence, the same request to the same
      endpoint might well be billed differently with
      regard to the status code (e.g., \$0.01 if 200 OKs and
      \$0.005 if 400 Bad Requests). We modeled this
      situation as a new metric for each status code. For
      example, in \textit{FacePlusPlus}, the \textit{QPS}
      metric has been split into \textit{QPS\_OK}, \textit{QPS\_timeout} and
      \textit{QPS\_invalidParam}.
    
    \textit{MI-03} If a certain plan explicitly denies access to certain API operations (e.g., \textit{Azure Search}), those operations are not included in the model.
    
    \textit{MI-04} If the actual value for a quota or
      rate is unknown (e.g., \textit{Accuweather}), we omit
      this rate/quota. For example, a number of APIs
      explicitly mention that they apply some rate-limiting
      value, but they do not mention what the actual value
      is.

    \textit{MI-05} Some metrics are dependent on some aspect
      of the HTTP request (body, parameters, etc.) and
      do not have any associated period (e.g.,
      \textit{FacePlusPlus}). In this case, the \textit{period}
      property is removed.

    \textit{MI-06} There are also pricings with unknown cost
      (such as educational plans, non-profit organizations,
      enterprise, etc.). These are modeled with \textit{custom: true} (e.g., \textit{GeoRanker}). Additionally, if a limitation has a custom value to be negotiated with its provider, it is also modeled with \textit{custom: true} (e.g., \textit{Yelp}).

    \textit{MI-07} In plans whose billing depends on
      the number of users (e.g., \textit{Box}) or on other variables affecting the cost
      (number of organizations, consumers, accounts,
      etc.), we considered the minimum number allowed. For example, plan \textit{Business Starter} of \textit{Box} requires a minimum of 3 users, and it becomes more expensive if more users join the plan. Therefore, for the sake of simplicity, we consider the cost of \textit{Business Starter} to be the cost for 3 users.
    
    \textit{MI-08} Finally, in APIs whose documentation does not specify whether the time window in which limits are calculated is fixed or sliding, we assumed that limits with longer periods (e.g. years or months) use fixed windows, and limits with shorter periods (e.g. seconds or minutes) use sliding windows. This decision is based on the research in \cite{Gamez-Diaz2017_icsoc_main}.

    \textbf{Open issues}:
    
    \textit{OI-01} Some HTTP query parameters are limited to a
      certain range of allowed values instead of a maximum
      value (e.g., \textit{Scopus}). Despite the fact that
      we  modeled some parameters as a metric (e.g.,
      number of results), parameters within a range were not modeled. In the Scopus case, \textit{Scopus
        Search} limits the number of results to 25 in the
      \textit{non-subscriber} plan, whereas this number
      rises to 200 in the \textit{subscriber} plan.
      Nevertheless, it also limits the parameter
      \textit{view} to \textit{STANDARD} in the former case and
      allows \textit{COMPLETE} only in the latter.

    \textit{OI-02} Another open issue arises when the overage
      cost is also limited (e.g., \textit{Georanker}). Some
      providers force one to move to another plan if one
      surpasses a certain value of the overage cost. This
      situation has not been modeled. For example, the
      \textit{small} plan includes 300\thinspace000
      requests, with an overage cost of 0.001\$ per request.
      However, this overage cost goes up to 750\thinspace000 requests.
      Once this amount is reached, one has to move to the
      \textit{medium} plan.

\subsection{RQ2 - Automation}
\label{sec:automatedAnalyzer}

The main operation regarding API limitations is
\textit{Validity} (cf.
Section~\ref{sec:analysis}). This operation, in order to
be useful for practitioners, needs to be automated by means of
a specific tool. To this end, we have developed
\textit{sla4oai-analyzer} \footnote{\url{https://github.com/isa-group/sla4oai-analyzer}} , a publicly available command-line tool prototype \cite{analyzer-repo}. Once installed, given a SLA4OAI file, the command \textit{sla4oai-analyzer -o <operation> -f <myFile.yaml>} will initiate the validity analysis for this file.
\begin{figure}[hbt!]
    \centering
    \includegraphics[width=1.\linewidth]{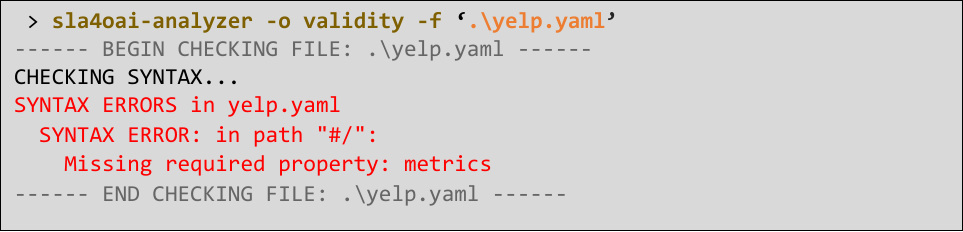}
    \caption{Tool running a syntax check.}
    \label{figure:tool-p0}
\end{figure}

\begin{figure}[hbt!]
    \centering
    \includegraphics[width=1.\linewidth]{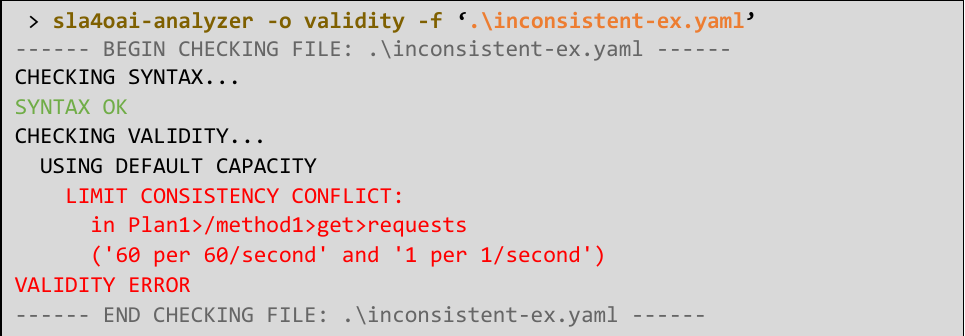}
    \caption{Tool running the validity operation with errors.}
    \label{figure:tool-p1-nok}
\end{figure}

For example, for the validity operation,
\textit{sla4oai-analyzer} first checks the syntax validity
according to the JSON Schema defined in the repository, and
then checks each validity criterion in each part (pricing,
plan, limitation, and limit). 
Figure~\ref{figure:tool-p1-nok} depicts a consistency
conflict detected by this tool, caused by a modeling mistake.

As illustrations of some outputs of the tool, Figure~\ref{figure:tool-p0} shows a pricing with
\textit{syntax errors} and 
Figure~\ref{figure:tool-p1-nok} a
\textit{consistency conflict}.

\begin{figure}[hbt!]
    \centering
    \includegraphics[width=1.\linewidth]{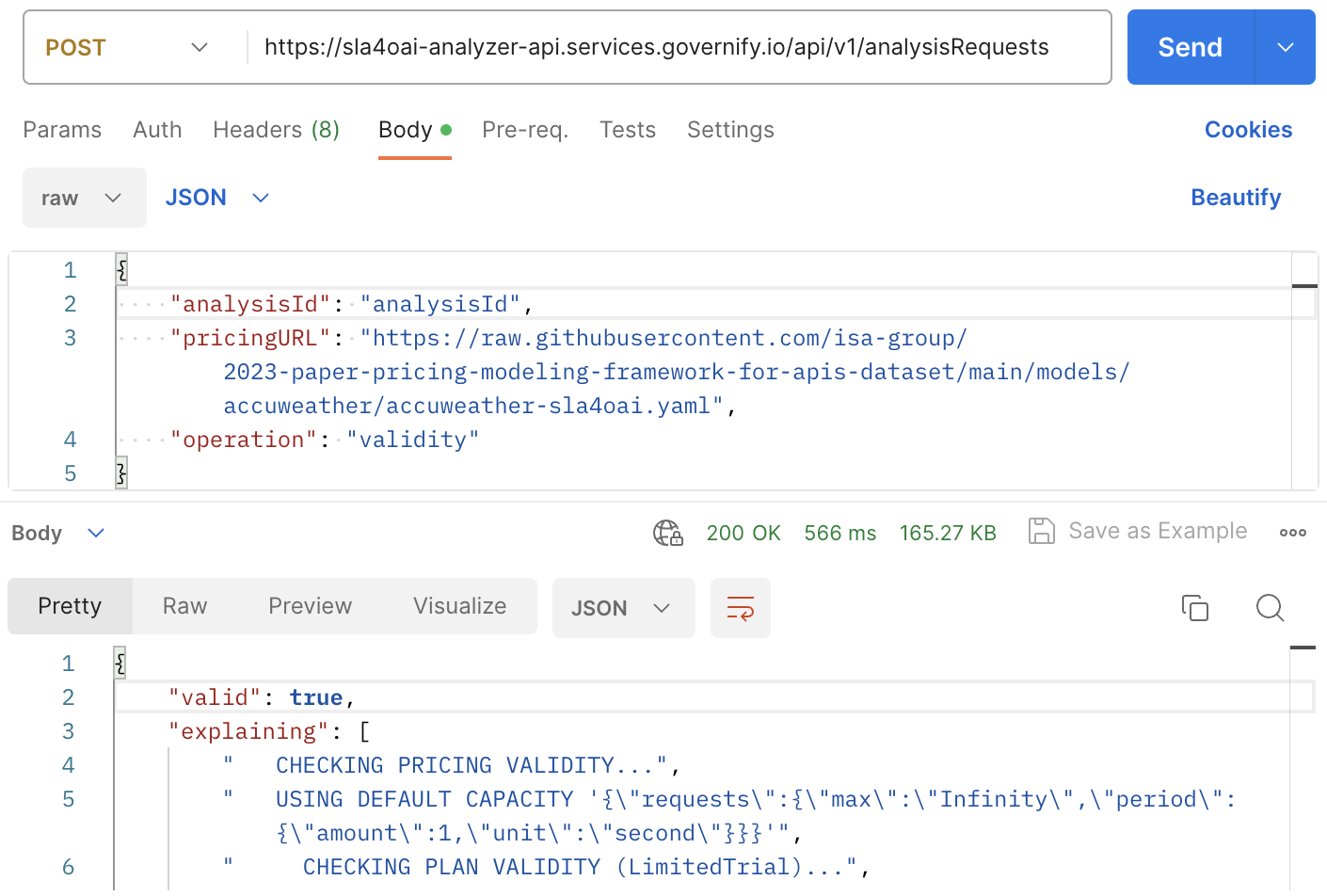}
    \caption{Simple UI for the \textit{sla4oai-analyzer} API.}
    \label{figure:sla4oaiAnalyzerAPI}
\end{figure}

This tool is also available as an API \cite{analyzer-api-repo}.
Furthermore, we provide a Postman documentation\footnote{\url{https://documenter.getpostman.com/view/683324/TVKEYHv8}} with 54 examples of invocations of the validity operation using this API. Figure~\ref{figure:sla4oaiAnalyzerAPI} is a screenshot of an invocation and response of the API analysing the validity of the \textit{Accuweather} pricing.

\subsection{Threats to validity}
We need to analyse the various validity threats that may have influenced our work, and the ways in which we tried to mitigate them.

\subsubsection{Internal validity}
These threats refer to the factors that introduce bias in our work and affect its utility. In our case, the main threat is the subjective and manual review process of the documentation of 32 different APIs. As a result, some limitations might have been overlooked and omitted in our models. To mitigate this threat, we checked each API multiple times and made the appropriate changes when necessary, recording each change, taking screenshots of their websites and saving their URLs. Moreover, some APIs updated their documentation over the span of time of writing this paper, so we updated their corresponding models. In some cases, pricing plans were removed from the API website, so we used the \textit{Wayback Machine} tool (\url{https://archive.org/web/}) to retrieve older versions of these pages.

\subsubsection{External validity}
This refers to the extent to which we can generalise from the results of our work. One threat is representativity. The APIs were extracted from
three sources --  217 valid APIs from the Neumann dataset,
2966 APIs replicating the extraction in Gamez-Diaz and 27 APIs from the RapidAPI most popular list. Since this
is too large a number to be analysed manually, we opted to
select a representative sample containing 54 APIs. This means that our model may not generalise to the rest of the APIs in the dataset. To mitigate this threat, we selected APIs from a wide range of domains, and some of them are popular and extensively consumed by a large number of users.

Another threat is that whereas our model supports the majority of the attributes
analysed, it does not support some of them. With that in
consideration, we concluded that we are able to model 85.3\%
(N=266) of the APIs regarding limitation attributes and
90.8\% (N=174) regarding pricing attributes, meaning that we are confident enough on the ability to generalise our model. 
Those APIs that can be modeled regarding both limitation
and pricing attributes comprise
78.2\% of the overall Pricing4APIs dataset.

Finally, our proposal has not yet been validated with other API consumers and providers. This means that SLA4OAI might not be as usable or useful as we intended. To mitigate this, we provide a JSON schema \cite{sla4oai-json-schema} to help understand the specification. Additionally, we proposed the addition of SLA4OAI as an official OpenAPI Specification extension.


\section{Related Work}
\label{sec:relatedWork}

In recent years, the information of real-world APIs has been analyzed from different perspectives. In~\cite{Neumann2018}, the authors  analyzed more than 500 publicly available APIs to
identify the different trends in the current industrial
landscape. In~\cite{Gamez-Diaz2017_icsoc_main}, we
analyzed a set of 69 real APIs in the industry to
characterize the variability of their offers, drawing a
number of invaluable conclusions about real-world APIs such
as: (i) most APIs provide different capabilities depending
on the tier or plan that the API consumer is willing to pay for;
(ii) usage limitations are a common aspect that all APIs describe
in their offers;  (iii) limitations on API requests are
the commonest, including quotas over static 
time periods (e.g., \textit{1000 requests each natural day}) and
rates for dynamic time periods (e.g., \textit{3 requests per
  second});  (iv) offers can include a broad number of
metrics over other aspects of the API that may be
domain-independent (such as the number of results returned
or the size of the request  in bytes) or domain-dependent
(such as the CPU/RAM consumption during the request
processing or the number of different resource types).
Based on these conclusions, we identified the need for
non-functional support in the API development life-cycle and
the high level of expressiveness present in the API
offers.

In this sense, the information of non-functional aspects has been studied in both industry and academia. In the industry, the term \textit{SLA} is usually related to elements such as availability or response time, commonly associated with \textit{guarantees} that may include compensations and penalties. In the academia, \textit{SLA} has been used with multiple meanings. In the literature, there are proposals that deal with SLAs in the context of web services \cite{Heiko2003}, while others deal with services in general \cite{Uriarte2014,Kouki2014}. Some proposals showcase the importance of having proper documentation \cite{shojaiemehr2019three}. Furthermore, there are open proposals under the term \textit{agreement}, such as WS-Agreement \cite{Andrieux2007} or L-USDL \cite{Garcia2017}. In any case, to the best of our knowledge, either \textit{SLA} or \textit{agreement} have been used to describe and model non-functional aspects of services (functional aspects such as operations or data types were not included in these proposals). In the following paragraphs we analyze the ability of those existing models to describe the information of a pricing (we will refer to them as \textit{SLA models} for the sake of simplicity). Specifically, in Table~\ref{table:SLAAnalysis}, we consider 6 aspects to analyze in each proposal based on various functional aspects of RESTful APIs such as multiple operations or hierarchical relationships.

\begin{table}[!ht]
    \centering
    \caption{Analysis of SLA models.}
    \label{table:SLAAnalysis}
    \begin{tabular}{@{}lllllllll@{}}
        \toprule
        \textbf{\textbf{Name}}                & \textbf{\textbf{F1}} & \textbf{\textbf{F2}} & \textbf{\textbf{F3}} & \textbf{\textbf{F4}} & \textbf{\textbf{F5}} & \textbf{\textbf{F6}} \\ \midrule
        \textbf{ysla}~\cite{Engel2018}   	  & YAML                 &                      & \checkmark               & \checkmark               & \checkmark               & \\
		\textbf{SLAC}~\cite{Uriarte2014}      & DSL                  &                      &                      &                      &                      & \\
        \textbf{CSLA}~\cite{Kouki2014}        & XML                  &                      & \checkmark               &                      &                      & \\
        \textbf{L-USDL}~\cite{Garcia2017} & RDF                  & \checkmark               & \checkmark               &                      & \ding{61}            &   \\
        \textbf{rSLA}~\cite{Tata2016}         & Ruby                 & \checkmark               &                      & \checkmark               & \checkmark               &   \\
        \textbf{SLAng}~\cite{Lamanna2003}     & XML                  & \checkmark               &                      &                      &                      &   \\
        \textbf{WSLA}~\cite{Heiko2003}        & XML                  & \checkmark               & \checkmark               &                      & \checkmark               &   \\
        \textbf{SLA*}~\cite{Kearney2010}      & XML                  & \checkmark               & \checkmark               &                      & \checkmark               &   \\
        \textbf{WS-Ag.}~\cite{Andrieux2007}   & XML                  & \checkmark               & \checkmark               & \checkmark               & \ding{61}            &   \\ 
        \textbf{Pricing4API} & YAML & \checkmark & \checkmark & \checkmark & \checkmark & \checkmark \\ \bottomrule
    \end{tabular}
    \bigskip
    \\
    \footnotesize\textit{\ding{61}} Supported with minor enhancements or modifications. \\
    \raggedright
    Comparative aspects:\\
\textbf{F1} The serialization format in which the document can be written.\\
\textbf{F2} Its target domain is web services.\\
\textbf{F3} It can model more than one operation.\\
\textbf{F4} It has a hierarchical models or overriding properties and metrics.\\
\textbf{F5} It can model temporal concerns.\\
\textbf{F6} It defines specific semantics for elements of REST API pricings.\\
    \vspace{-2ex}
\end{table}

Based on the comparison of the different SLA models, we
would draw the following conclusions. (i) None of the
specifications provides any support for or alignment with
the OpenAPI Specification. (ii) Most of the approaches
provide a specific syntax for RDF/XML (\textbf{F1}) (some, however, lack
such a syntax), but there is no explicit support for YAML or
JSON serializations, which are preferable in order to align with the OpenAPI Specification and also because these formats are processed and parsed faster using modern technological stacks \cite{nurseitov2009comparison}. (iii) While many proposals are
complete, others leave some parts open for practitioners to
implement. (iv) A number of proposals aim to model Web
services (\textbf{F2}) but are focused on traditional SOAP Web services
rather than RESTful APIs, so that they do not address the
modeling standardization of the RESTful approach in which
the concept of a resource is clearly unified (a URL), and
the amount of operations is limited (to HTTP methods such as
GET, POST, PUT and DELETE). This lack of support for RESTful
modeling and its semantics prevents the approach from having a concise and
compact binding between its functional and non-functional
aspects. (v) They have insufficient expressiveness to model
such limitations as quotas and rates for each resource and
method, and with full management of the temporality (static
or sliding time windows and periodicity) present in the
typical industrial API pricings (\textbf{F6}). (vi) Some are designed to
model a single operations (\textbf{F3}), and usually lack support for
hierarchical modeling or overriding properties and metrics
(\textbf{F4}). In such a context, they cannot model a set of tiers
or plans that yield a complex offer which maintains
coherence. Instead, they rely on a manual process to maintain the coherence between a set of models, opening the door to inconsistencies.

Existing notions of \textit{validity} in agreements and SLAs do not take into account the conflicts that appear due to inconsistencies with rates, quotas or prices. Our proposal completes and complements the notion from existing work on WS-Agreement \cite{Andrieux2007} and its extensions, focusing on the validation and compliance of the specific features of API pricings that were not discussed before.
 
There has also been a  proposal of a model-driven
approach to defining API service licenses and an API SLA
analyzer system which utilizes the proposed license model to
uncover SLA violations in real-time~\cite{Vukovic2014}.
Other work has sought to determine whether one can know how to price
SaaS by summarizing existing knowledge from different
research areas and SaaS pricing practice~\cite{Saltan2019}, and~\cite{Yrjonkoski}
presents a three-level productization model for different
phases of SaaS businesses.

To the best of our knowledge, while there does exist previous work on analyzing pricings in general, there has been no work focused on relating API pricings and limitations (i.e., quotas, rates).

We have some previous studies in this field, but they have significant differences with this paper. In \cite{Gamez-Diaz2018_icsoc_demo}, we presented a proof of concept implementation of a tool to highlight the importance of the automated analysis of induced limitations in limitation-aware microservices architectures, but there was no standard model for the pricings; nonetheless, this serves as a motivational use case.
In \cite{Gamez-Diaz2019_esecfse_demo}, we introduced an ecosystem of tools to support SLAs in APIs named \textit{Governify for APIs}, focusing on the development of limitation-aware API clients; however, there were no details about the definition of API pricings.
In \cite{Gamez-Diaz2019_esecfse_industry}, we conducted a survey to ascertain the importance of the role of SLAs in the development lifecycle of SLA-driven APIs in an industrial context; this serves as a motivation for this paper, but no further details about how to model the pricings were given at the time.
In \cite{Gamez-Diaz2019_icsoc_main}, we presented the syntax of an initial proposal of SLA4OAI, with a small modeling sample; this proposal was limited and lacked several elements that are now introduced in this paper (such as overage costs and globbing) that are key features in order to model real-world pricings.

\section{Future Work}
\label{sec:futureWork}

There are various open issues and known limitations that we would like to tackle in the future. In \cite{CloudComputingPatterns2014}, the authors distinguish five types of incoming workloads:  \textit{static}, \textit{periodic}, \textit{once-in-a-lifetime}, \textit{unpredictable}, and \textit{continuously changing}.
If we introduce the concept of temporality in the pricing, i.e., to consider that certain plans have a determined temporal validity (e.g., day/night plan), the operations have to be adapted to consider this temporality.
Joining temporality with workload models, one could automate the management of this type of advanced scenarios which require infrastructures that are dynamic (e.g., instances that start or stop and have a variable cost).

Furthermore, alert systems can be defined which
notify users  when certain
percentages of consumption of the limitations are reached, so that they can take this situation into
account and adjust their consumption accordingly.

With respect to the tool, as it is just a proof of concept,
it lacks various features. (i) It is a command-line tool and
an API, and is not really useful for end-users. (ii) The
implementation of \textit{cost conflicts} is too naive
as it only supports simple cost values (but neither overages nor
operation costs).

In our model, we identified two open issues that should be
addressed: (i) extend the model to incorporate parameters
that are limited to a certain range of allowed values
instead of a maximum value; (ii) expand the overage concept to
establish a limit on the overage itself.


\section{Conclusions}
\label{sec:conclusions}

In this paper, we have proposed the \textit{Pricing4APIs} model that
structure the nature of pricing plans and limitations present in RESTful API.
To design this model, we have analyzed both the limitations and the
pricing plans of a set of 268 real-world APIs using three different
datasets.  We then presented a curated list
of 54 API pricings with a formal validated model that
represents the variability found in the industry.  The
Pricing4APIs dataset used in the analysis is publicly
available as it could be a useful resource for both academics
and practitioners.

Concerning SLA4OAI (the proposed serialization of the Pricing4APIs model), it is important to note that, while it was designed to be aligned with the OpenAPI Specification, it could be adapted with minimal changes to other API descriptions (such as RAML, API Blueprint, I/O Docs, WSDL, or WADL). However, given the vibrant community that embraces OpenAPI, we believe that the proposed SLA4OAI can pave the way to creating an open ecosystem of tools to automate the development process that takes into account the cost and limitation information: from frameworks to predict or test the estimated capacity of APIs, to tools that automate the configuration of API gateways/proxies or throttling.  In this context, as a first step in this direction, in this paper we have also presented a validation tool that automates the analysis of pricings and spot for inconsistencies or conflicts.

    \section*{Acknowledgments}

This work has been partially supported by the grant PID2021-126227NB-C22 funded by MCIN /AEI /10.13039 /501100011033 /FEDER,UE; TED2021-131023B-C21 funded by MCIN /AEI /10.13039 /501100011033 and by European Union ``NextGenerationEU'' /PRTR; and the FPU scholarships FPU15/02980 and FPU19/00666 funded by the Spanish Ministry of Universities. The authors wish to thank Daniel Arteaga and Felipe Vieira da Cunha for their invaluable technical contributions.



\bibliographystyle{elsarticle-num} 
\bibliography{references}

%


\vspace{5mm}

\begin{wrapfigure}{l}{25mm}
\includegraphics[width=1in,height=1.25in,clip,keepaspectratio]{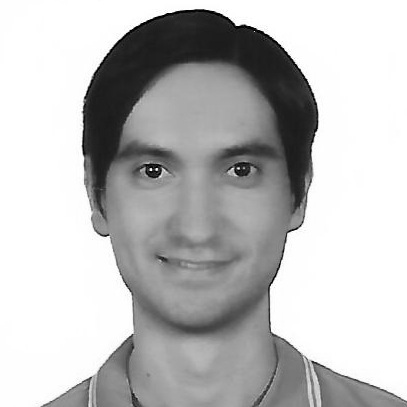}
\end{wrapfigure} \par
\noindent \textbf{Rafael Fresno-Aranda} is a predoctoral researcher at the
  University of Sevilla. He received a BSc degree in 2019 and
  an MSc degree in Software Engineering in 2020. He recently got a
  competitive predoctoral fellowship (FPU), granted by the
  Spanish government. His research focuses on
  Service-Oriented Computing, including the analysis of
  microservices architectures and RESTful APIs. He has contributed
  to open source tools and utilities, and has collaborated
  with the University of California, Berkeley to develop an auditor
  framework for software engineering students.
  Contact him at rfresno@us.es.

\vspace{5mm}

\begin{wrapfigure}{l}{25mm}
\includegraphics[width=1in,height=1.25in,clip,keepaspectratio]{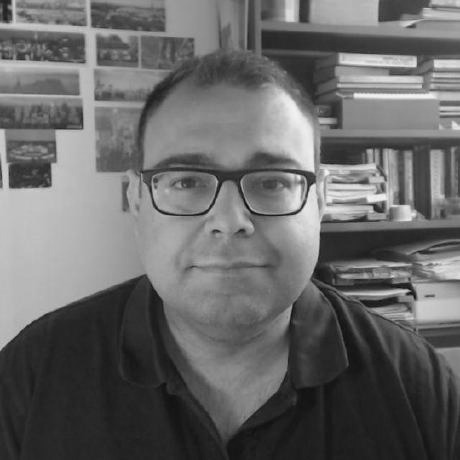}
\end{wrapfigure} \par
\noindent \textbf{Pablo Fernández} is an associate professor at the University
  of Sevilla, Spain, and a member of the ISA Research Group.
  His current research is focused on the automated
  governance of organizations based on SLAs and commitments. He has been the lead architect
  for several PPP projects in scenarios of public
  administrations and major firms. Contact him at
  pablofm@us.es;
  \href{https://www.isa.us.es/members/pablo.fernandez}{https://www.isa.us.es/members/pablo.fernandez}

\vspace{5mm}

\begin{wrapfigure}{l}{25mm}
\includegraphics[width=1in,height=1.25in,clip,keepaspectratio]{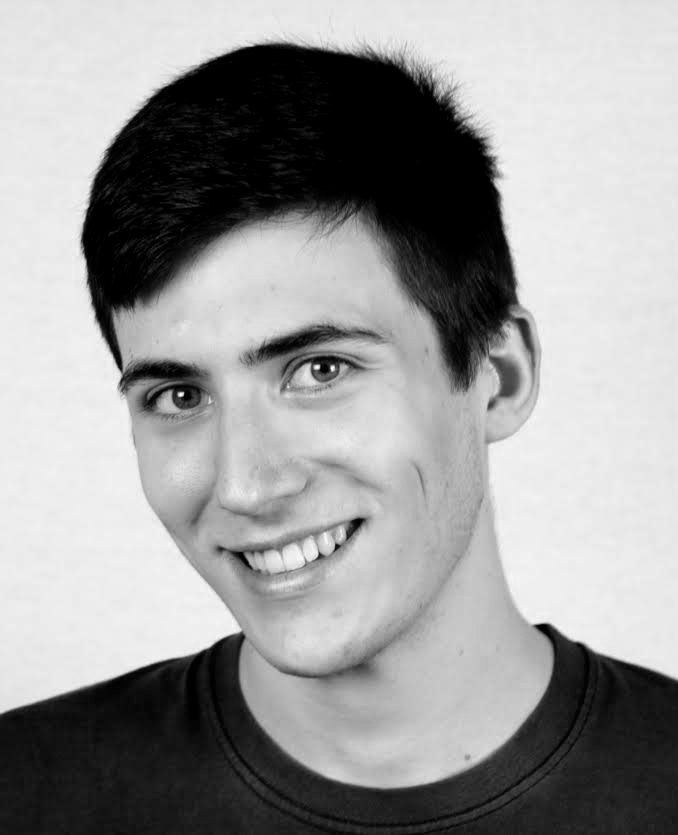}
\end{wrapfigure} \par
\noindent \textbf{Antonio
    Gámez-Díaz}, PhD in Software Engineering from the Universidad de Sevilla (2022), where he received his BSc (2015) and MSc (2016) degrees. He got a competitive predoctoral fellowship (FPU), granted by the Spanish government.
With a passion for teaching and collaborating with industry leaders, he remains committed to his research interests in Service-Oriented Computing. However, due to the precarious situation of science in his country, he left
the Academia in 2021 and is working at VMware, where he contributes to a number of cloud-native open-source projects.
  Contact him at antoniogamez@us.es;
  \href{https://personal.us.es/agamez2}{https://personal.us.es/agamez2}.

\vspace{5mm}

\begin{wrapfigure}{l}{25mm}
\includegraphics[width=1in,height=1.25in,clip,keepaspectratio]{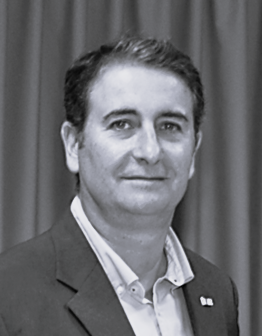}
\end{wrapfigure} \par
\textbf{Prof. Dr Amador Duran} is an associate professor of Software Engineering at the University of Seville, Spain, and a member of the ISA Research Group. His current research focuses on empirical software engineering, requirements engineering, and metamorphic testing.  
He is the author of the \href{http://www.lsi.us.es/descargas/descarga_programas.php?id=3&lang=en}{\textsc{REM} tool}, used by universities and companies in various countries. He also serves regularly as a reviewer for international journals and conferences.
Contact him at amador@us.es; \href{https://www.isa.us.es/members/amador.duran}{https://www.isa.us.es/members/amador.duran}.

\vspace{5mm}

\begin{wrapfigure}{l}{25mm}
\includegraphics[width=1in,height=1.25in,clip,keepaspectratio]{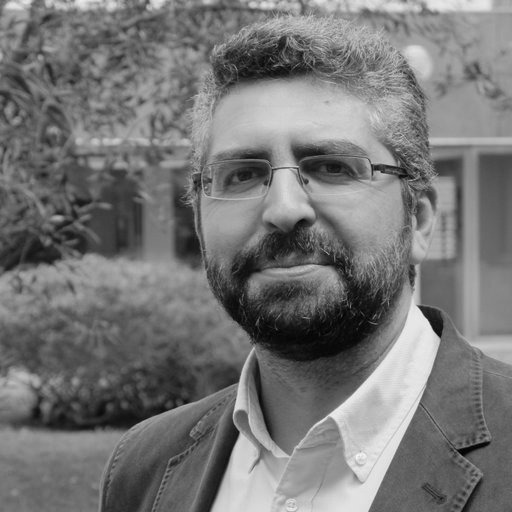}
\end{wrapfigure} \par
\textbf{Prof. Dr Antonio Ruiz-Cortés} is a Full Professor and head of
  the Applied Software Engineering Group at the University
  of Seville (Spain). His current research focuses on
  service-oriented computing, business process management,
  testing and software product lines. He is an associate
  editor of Springer Computing, recipient of the Most
  Influential Paper of SPLC (2017) and VAMOS award (2020),
  and elected member of the Academy of Europe.  Contact him
  at aruiz@us.es;
  \href{https://www.isa.us.es/members/antonio.ruiz}{https://www.isa.us.es/members/antonio.ruiz}.




%
%
%
%
\end{document}